\documentclass[thmsa,12pt]{article}

\usepackage{geometry}

\usepackage{graphicx}
\usepackage{enumerate}
\usepackage{courier}
\usepackage{times}
\usepackage{amsmath,amssymb}

\usepackage{hyperref}

\newtheorem{Theorem}{Theorem}[section]
\newtheorem{Proposition}[Theorem]{Proposition}

\newtheorem{Corollary}[Theorem]{Corollary}
\newtheorem{Lemma}[Theorem]{Lemma}

\newenvironment{Proof}
{\begin{trivlist}\item[]{{\sc Proof.}}}{\hfill{$\square$}\noindent\end{trivlist}}

\newtheorem{Definition}[Theorem]{Definition}

\newtheorem{Example}[Theorem]{Example}

\newtheorem{Remark}[Theorem]{Remark}

\newcommand{\cM}{\mathcal{M}}

\newcommand{\cH}{\mathcal{H}}
\newcommand{\cP}{\mathcal{P}}
\newcommand{\cS}{\mathcal{S}}
\newcommand{\cT}{\mathcal{T}}
\newcommand{\PG}{\operatorname{PG}}
\newcommand{\N}{\mathbb{N}}

\newcommand{\F}{\mathbb{F}}
\newcommand{\wt}{\operatorname{wt}_{\text{H}}}
\newcommand{\we}{\operatorname{w}_C^{\text{H}}}
\newcommand{\web}{\operatorname{w}_C^{b}}
\newcommand{\dham}{\operatorname{d}_{\text{H}}}
\newcommand{\db}{\operatorname{d}_{b}}
\newcommand{\wtb}{\operatorname{wt}_{b}}

\begin{document}


\title{Linear codes for $b$-symbol read channels attaining the Griesmer bound}

\author{Sascha Kurz\\ University of Bayreuth\\ sascha.kurz@uni-bayreuth.de}

\date{}

\maketitle

\noindent
\textbf{Abstract} Reading channels where $b$-tuples of adjacent symbols are read at every step have e.g.\ applications in storage. Corresponding bounds 
and constructions of codes for the $b$-symbol metric, especially the pair-symbol metric where $b=2$, were intensively studied in the last fifteen years. 
Here we determine the optimal code parameters of linear codes in the $b$-symbol metric assuming that the minimum distance is sufficiently large. We also 
determine the optimal parameters of linear binary codes in the pair-symbol metric for small dimensions.

\smallskip


\section{Introduction}
In storage applications the reading device is sometimes insufficient to isolate adjacent symbols, which makes it necessary to adjust the standard coding-theoretic 
error model. Cassuto and Blaum studied a model where pairs of adjacent symbols are read in every step and introduced the so-called symbol-pair metric for codes 
\cite{cassuto2011codes}. This notion was generalized to the $b$-symbol metric where $b$-tuples of adjacent symbols are read at every step, see e.g.\ \cite{yaakobi2016constructions}. 
While general codes where studied, see e.g.\ \cite{chen2022new}, for representation and decoding purposes it is beneficial to assume more structured codes. A rather general 
and important subclass of codes are linear codes i.e.\ subspaces of some vector space $\F_q^k$ over a finite field. For small minimum distances a Singleton type bound, 
introduced by Chee et al.\ for the symbol-pair metric \cite{chee2013maximum}, turned out to be very effective. Like for the Hamming metric codes attaining this bound are called 
maximum distance separable (MDS) codes and quite some constructions were studied for the symbol-pair metric, see e.g.\ \cite{ding2018new,kai2015construction,kai2018new,luo2022new,
ma2021symbol,ma2022constructions,ma2022mds}. Here we are interested in optimal linear codes w.r.t.\ the $b$-symbol metric in those situations where the minimum distance is \emph{large}. 
For the Hamming metric Solomon and Stiffler \cite{solomon1965algebraically} showed that the so-called \emph{Griesmer bound} \cite{griesmer1960bound} can always be attained with equality 
if the minimum distance is sufficiently large. The main objective of this paper is to show the analogous result for the $b$-symbol metric for the Griesmer type bound recently 
introduced in \cite{huang2025b,luo2024griesmer}. In order to get a more complete picture of the intermediate situation we determine the parameters of optimal binary linear codes in the symbol-pair 
metric for small dimensions.

The remaining part of the paper is structured as follows. In Section~\ref{sec_preliminaries} we introduce the necessary preliminaries. Our main result on the minimum length 
of a linear code in the $b$-symbol metric for large minimum distances in stated in Theorem~\ref{thm_main} in Section~\ref{sec_constructions}. To complement this asymptotic result we 
determine the exact minimum lengths of binary linear codes in the $b$-symbol metric for small dimensions in Section~\ref{sec_exact_values}. The paper is closed by a brief conclusion
in Section~\ref{sec_conclusion}. 

\section{Preliminaries}
\label{sec_preliminaries}
For some prime power $q$ let $\F_q$ denote the finite field with $q$ elements and $\F_q^n$ the $n$-dimensional vector space over $\F_q$. An \emph{$[n,k]_q$ code} $C$ is a $k$-dimensional 
linear subspace of $\F_q^n$. If $C$ is given as the row span of a $k\times n$ matrix $G$, then $G$ is called a \emph{generator matrix} of $C$. A generator matrix $G$ is called systematic 
if it starts with a $k\times k$ identity matrix. The elements of an $[n,k]_q$ code $C$ are called \emph{codewords}. The \emph{Hamming weight} $\operatorname{wt}_{\text{H}}(c)$ of a codeword 
$c=\left(c_0,\cdots,c_{n-1}\right)$ is the number of non-zero entries $\left| \left\{c_i\,:\, c_i\neq 0, 0\le i\le n-1\right\}\right|$. With this, the \emph{weight enumerator} of $C$ is 
the homogeneous polynomial $\we(x,y):=\sum_{c\in C} x^{\wt(c)}y^{n-\wt(c)}$ and the \emph{Hamming distance} between two codewords $c,c'\in C$ is given by $\dham(c,c'):=\wt(c-c')$.  

The Hamming distance is an appropriate measure for error detection and correction in certain channels and is indeed a metric. In \cite{cassuto2011codes} the authors introduced another 
channel where a different kind of metric is more suitable. Instead of single symbols one assumes that neighboring pairs of symbols are read, which was subsequently generalized to 
neighboring $b$-tuples of symbols for $b\ge 2$, see e.g.\ \cite{yaakobi2016constructions} for details. To this end let 
\begin{equation}
  \label{eq_read_vector}
  \pi_b(c):=\big((c_0,\dots,c_{b-1}),(c_1,\dots,c_b),\dots,(c_{n-1},c_{0},\dots,c_{b-2})\big)
\end{equation} 
be the \emph{$b$-symbol read vector} of a codeword $c=(c_0,\dots,c_{n-1})\in C\subseteq \F_q^n$ and
\begin{equation}
  \label{eq_dist}
  \db(c,c'):=\left|\left\{0\le i\le n-1\,:\, \left(c_i,c_{i+1},\dots,c_{i+b-1}\right)\neq\left(c_i',c_{i+1}',\dots,c_{i+b-1}'\right)\right\}\right|  
\end{equation}
be the \emph{$b$-symbol distance} between codewords $c=\left(c_0,\dots,c_{n-1}\right)$ and $c'=\left(c_0',\dots,c_{n-1}'\right)$, where the indices are read modulo $n$. Similar to the 
definition of $\wt(c)$ we can define the \emph{$b$-weight} $\wtb(c):=\db(c,0)$ of a codeword $c\in C$ and the corresponding weight enumerator 
$\web(x,y):=\sum_{c\in C} x^{\wtb(c)}y^{n-\wtb(c)}$ of $C$ (w.r.t.~the $b$-symbol distance). We remark that $\operatorname{d}_1$ is equal to the Hamming distance and 
$\operatorname{d}_2$ denotes the pair-symbol distance.

The \emph{minimum Hamming distance} $\dham(C)$ of an $[n,k]_q$ code $C$ is defined as the minimum Hamming distance between two different codewords, i.e.\ 
\begin{equation}
  \dham(C):=\min\!\left\{\dham(c,c')\,:\,c,c'\in C, c\neq c'\right\}=\min\!\left\{\wt(c)\,:\, c\in C, c\neq 0\right\}.
\end{equation}   
If $d=\dham(C)$ then we also speak of an \emph{$[n,k,d]_q$ code}. Similarly, 
\begin{equation}
  \db(C):=\min\!\left\{\db(c,c')\,:\,c,c'\in C, c\neq c'\right\}=\min\!\left\{\wtb(c)\,:\, c\in C, c\neq 0\right\}
\end{equation}  
and we speak of an \emph{$[n,k,d]_q^{b}$ code} if $d=\db(C)$. We call an $[n,k,d]_q$ code \emph{optimal} (w.r.t.\ the 
Hamming metric) if no $[n-1,k,d]_q$ code exists. Similarly, we call an $[n,k,d]_q^{b}$ code \emph{optimal} (w.r.t.\ the $b$-symbol metric) 
if no $[n-1,k,d]_q^{b}$ code exists. Note that there are several notions of optimality for linear codes and here we choose length-optimality, 
i.e.\ the smallest possible length for given parameters $k$, $d$, $q$, and $b$, which is justified by the following observation. 

\begin{Lemma}
  The existence of an $[n,k,d]_q^{b}$ code implies the existence of an $[n+1,k,\ge d]_q^{b}$ code. 
\end{Lemma}
\begin{Proof}
  If $G$ is the generator matrix of an $[n,k,d]_q^{b}$ code, then appending a zero vector to $G$ yields a generator matrix of an $[n+1,k,\ge d]_q^{b}$ code. 
\end{Proof}
So, by $n_q(k,d)$ we denote the smallest integer $n$ such that an $[n,k,d]_q$ code exists and by $n_q^{b}(k,d)$ we denote the smallest integer $n$ such that an $[n,k,d]_q^{b}$ code exists. 
While the determination of $n_q(k,d)$, for certain parameters, is a classical problem in coding theory, besides some general upper and lower bounds for $n_q^{b}(k,d)$, not many exact 
values of $n_q^{b}(k,d)$ are known. So, we aim to partially close this gap to determining $n_2^{2}(k,d)$ for small values of the dimension $k$, see Section~\ref{sec_exact_values}. For 
linear codes w.r.t.\ the Hamming metric e.g.\ the so-called
\emph{Griesmer bound} \cite{griesmer1960bound}
\begin{equation}
  \label{eq_griesmer_bound}
  n\ge \sum_{i=0}^{k-1} \left\lceil\frac{d}{q^i}\right\rceil=:g_q(k,d)
\end{equation}   
relates the parameters of an $[n,k,d]_q$ code. Interestingly enough, this bound can always be attained with equality if the minimum distance $d$ is sufficiently large 
and a nice geometric construction was given by Solomon and Stiffler \cite{solomon1965algebraically}.\footnote{More precisely, the cited papers show the statements for field 
size $q=2$, while they were generalized by other authors slightly later.} In other words, we have $n_q(k,d)=g_q(k,d)$ for all sufficiently large $d$ given $k$ and $q$. While this solves the 
asymptotic case, the full determination of the function $n_q(k,\cdot)$ is still a rather challenging problem that is solved in the binary case $q=2$ for dimensions $k\le 8$ 
only \cite{bouyukhev2000smallest}. It is known that $n_q(k,d)=g_q(k,d)$ for $k\le 4$ when $q=2$ and for $k\le 2$ where $q$ is an arbitrary prime power. In all other cases we 
only know $n_q(k,d)\ge g_q(k,d)$ for all $d\in\mathbb{N}$ and that there exists at least one, but only finitely many, $d\in\mathbb{N}$ such that $n_q(k,d)> g_q(k,d)$. 

From the many bounds for $[n,k,d]_q^{b}$ codes from the literature we would like to single out
\begin{equation} 
  \label{eq_griesmer_bound_pair_distance}
  \frac{q^b-1}{q-1}\cdot n \ge g_q\!\left(k,q^{b-1}\cdot d\right),
\end{equation}
see \cite[Theorem 1]{luo2024griesmer}. More precisely, starting from an $[n,k]_q^b$ code $C$ with minimum distance $d$ and a generator matrix $G$ of a 
$\left[\tfrac{q^b-1}{q-1},b,q^{b-1}\right]_q$ simplex code an $\left[\tfrac{q^b-1}{q-1}\cdot n,k\right]_q$ code $C'$ 
is constructed as follows. For each codeword $c\in C$ multiply $G$ with all elements of $\pi_b(c)$ and concatenate the results to a codeword $c'\in C$, so that 
$\wt(c')=\wtb(c)\cdot q^{b-1}$, since all non-zero codewords of a $b$-dimensional simplex code over $\F_q$ have Hamming weight $q^{b-1}$. Applying 
the Griesmer bound to $C'$ gives Inequality~(\ref{eq_griesmer_bound_pair_distance}). Given an $[n,k,d]_q^b$ code $C$ we call $C'$ the \emph{associated Hamming code} 
and will state some of its basic properties. To this end let us call an $[n,k]_q$ code or an $[n,k]_q^b$ code \emph{$\Delta$-divisible} if the weight of every 
codeword $c$, i.e.\ $\wt(c)$ or $\wtb(c)$, is divisible by $\Delta$.
\begin{Lemma}
  \label{lemma_code_properties}
  Let $C'$ be the Hamming code associated to an $[n,k,d]_q^b$ code $C$. Then, $C$ is a $q^{b-1}$-divisible $\left[\tfrac{q^b-1}{q-1}\cdot n,k,q^{b-1}\cdot d\right]_q$ code 
  with maximum weight at most $q^{b-1}\cdot n$. 
\end{Lemma}
\begin{Proof}
  Given the above reasoning it suffices to observe that the maximum weight of $C$ is at most $n$.
\end{Proof}
\begin{Example}
  For $n=5$, $q=2$, and $b=2$ let $c_1=(10101)$, $c_2=(11110)$, and $c_3=(11100)$ so that the read vectors are given by $\pi_2(c_1)=\big((1,\!0),(0,\!1),(1,\!0),(0,\!1),(1,\!1)\big)$,  
  $\pi_2(c_2)=\big((1,\!1),(1,\!1),(1,\!1),(1,\!0),(0,\!1)\big)$, and $\pi_2(c_3)=\big((1,\!1),(1,\!1),(1,\!0),(0,\!0),(0,\!1)\big)$. Using the generator matrix 
  $\left(\begin{smallmatrix}1&1&0\\0&1&1\end{smallmatrix}\right)$ for the $[3,2,2]_2$ simplex code,  
  $c_1$ is mapped to $c_1'=(110\,011\,110\,011\,101)$, $c_2$ is mapped to $c_2'=(101\,101\,101\,110\,011)$, and $c_3$ is mapped to $c_3'=(101\,101\,110\,000\,011)$. 
  Setting $C_1:=\langle c_1\rangle$, $C_2:=\langle c_2\rangle$, and $C_3:=\langle c_3\rangle$ we have that the $C_i$ are $[5,1]_2$ codes with weight enumerators $y^5 x^0+x^3y^2$, 
  $y^5 x^0+x^4y^1$, and $y^5 x^0+x^3y^2$, respectively. Moreover, we have $\operatorname{d}_2(C_1)=\wt(c_1')/2=5$, $\operatorname{d}_2(C_2)=\wt(c_2')/2=5$, and 
  $\operatorname{d}_2(C_3)=\wt(c_3')/2=4$.   
\end{Example}

Clearly the minimum $b$-symbol distance $\db(C)$ of an $[n,k]_q$ code $C$ can be bounded by its minimum Hamming distance as follows:
\begin{equation} 
  \min\{\dham(C)+b-1,n\}\le \db(C)\le \min\{b\cdot \dham(C),n\}.
\end{equation}
For linear cyclic codes and $b=2$ the lower bound was improved to $\operatorname{d}_2(C)\ge \left\lceil\tfrac{3\cdot\dham(C)}{2}\right\rceil$ \cite[Lemma 2]{yaakobi2016constructions}.

\begin{Remark}
  Given an $[n,k,d]_q^{b}$ code $C$ we can construct a $\left[\tfrac{q^b-1}{q-1}\cdot n,k,q^{b-1}\cdot d\right]_q$ code $C'$. The other direction is not always possible. As an example consider the 
  generator matrix
  $$
    \begin{pmatrix}
      101 & 000 & 101 & 011 & 110 \\
      011 & 000 & 011 & 110 & 101 \\
      000 & 101 & 101 & 101 & 101 \\
      000 & 011 & 011 & 011 & 011 \\
    \end{pmatrix}
  $$
  that generates a $[15,4,8]_2$ code $C'$. Here we have grouped the columns in pairs of three, which geometrically corresponds to a so called line spread of $\PG(3,2)$. Indeed we can 
  compute corresponding $2$-symbol read vectors from the four generating codewords
  $$
    \begin{pmatrix}
      10 & 00 & 10 & 01 & 11 \\
      01 & 00 & 01 & 11 & 10 \\  
      00 & 10 & 10 & 10 & 10 \\
      00 & 01 & 01 & 01 & 01
    \end{pmatrix}\!.
  $$ 
  However, in none of the rows we can find an element $c\in \F_2^5$ such that $\pi_2(c)$ would equal the corresponding row. While we have some freedom in permuting the columns of a generator
  matrix of a linear code without changing its weight enumerator w.r.t.\ the Hamming metric we cannot end up with a $[5,4,4]_2^{2}$ code since such a code does not exist. To this end we
  note that computing the reduced echelon form of the generator matrix of a code does not change the code and that the possible generator matrices in reduced form echelon are given by
  $$
    \left(\begin{smallmatrix}
    1000*\\
    0100*\\
    0010*\\
    0001*
    \end{smallmatrix}\right)\!,
    \left(\begin{smallmatrix}
    100*0\\
    010*0\\
    001*0\\
    00001
    \end{smallmatrix}\right)\!,
    \left(\begin{smallmatrix}
    10*00\\
    01*00\\
    00010\\
    00001
    \end{smallmatrix}\right)\!,
    \left(\begin{smallmatrix}
    1*000\\
    00100\\
    00010\\
    00001
    \end{smallmatrix}\right)\!,
    \left(\begin{smallmatrix}
    01000\\
    00100\\
    00010\\
    00001
    \end{smallmatrix}\right).
  $$
  Since the fourth row always starts with at least three consecutive zeroes the minimum symbol-pair distance is at most $5-2=3$ and not $4$.
\end{Remark}

\medskip

It is well known that linear codes correspond to multisets of points in projective geometries \cite{dodunekov1998codes}.  
The set of all subspaces of $\F_q^r$ , ordered by the incidence relation
$\subseteq$, is called \emph{$(r-1)$-dimensional projective geometry over
$\F_q$} and denoted by $\PG(r-1,q)$. Employing this algebraic notion
of dimension instead of the geometric one, we will use the term $i$-space
to denote an $i$-dimensional subspace of $\F_q^r$. To highlight the
important geometric interpretation of subspaces we will call $1$-, $2-$,
and $(r-1)$-spaces points, lines, and hyperplanes, respectively. For two
subspaces $S$ and $S'$ we write $S\le S'$ if $S$ is contained in $S'$. Moreover,
we say that $S$ and $S'$ are \emph{incident} iff $S\le S'$ or $S\ge S'$.
Let $[i]_q:=\tfrac{q^i-1}{q-1}$ denote the number of points
of an arbitrary $i$-space in $\PG(r-1,q)$ where $r\ge i$.
 Here we describe a multiset of points by a mapping $\mathcal{M}$ from the set of 
points of $\PG(k-1,q)$ to $\mathbb{N}$ and call $\mathcal{M}(P)$ the multiplicity of point $P$. The \emph{cardinality} of $\cM$ is given by $\#\cM=\sum_P \cM(P)$.  
For some subspace $S$ we define the multiplicity of $S$ by 
$\mathcal{M}(S)=\sum_{P\le S} \mathcal{M}(P)$ and let $\chi_S$ denote the characteristic function of $S$, i.e., $\chi_S(P)=1$ if $P\le S$ and $\chi_S(P)=0$ otherwise.
A multiset of points is called \emph{spanning} if the set of points with positive multiplicity span the entire space. With this, we can state more precisely
that each $[n,k,d]_q$ code with dual minimum distance at least two, i.e.\ for each coordinate there exists a codeword with non-zero entry at this position, is in 
one-to-one correspondence to a spanning multiset $\mathcal{M}$ in $\PG(k-1,q)$ with cardinality $n$ and $\mathcal{M}(H)\le n-d$ for every hyperplane $H$, where equality
occurs at least once.

Multisets of points can be generalized as follows, see \cite[Definition 4]{ball2025griesmer} and \cite[Definition 1]{kurz2024additive}.
\begin{Definition}
  \label{def_system}
  A \emph{projective $h-(n,r,s)_q$ system} is a multiset $\cS$ of $n$ subspaces
  of $\PG(r-1, q)$ of dimension at most $h$ such that each hyperplane
  contains at most $s$ elements of $\cS$ and some hyperplane contains
  exactly $s$ elements of $\cS$. We say that $\cS$ is \emph{faithful} if all of its 
  elements have dimension $h$. A projective $h-(n,r,s)_q$ system $\cS$
  is a projective \emph{$h-(n,r,s,\mu)_q$ system} if each point is contained
  in at most $\mu$ elements from $\cS$ and there is some point that is
  contained in exactly  $\mu$ elements from $\cS$.
\end{Definition}

Allowing $0$-spaces, corresponding to zero-columns in the generator matrix of a linear code, one can say that $[n,k,d]_q$ codes are in one-to-one correspondence to
projective $1-(n,k,n-d)_q$ systems. In general, projective $h-(n,r,s)_q$ systems (with $s<n$) are in one-to-one correspondence to additive codes, see \cite{ball2025griesmer} 
or \cite{kurz2024additive} for details. Here, we call a subset of $\mathbb{F}_q^n$ an \emph{additive code} iff the sum of any two codewords is also a codeword, i.e.\ additive codes 
are a super class of linear codes.  

\begin{Lemma}
  \label{lemma_associated_projective_system}
  To each $[n,k,d]_q^b$ code $C$, where $k\ge b$, we can associate a projective $b-(n,k,n-d)_q$ system $\cS=\left\{S_0,\dots,S_{n-1}\right\}$. Moreover, we have 
  \begin{equation}
    \label{eq_dim_intersection}
    \dim\!\left(S_i\cap S_{i+1}\right)\ge \max\!\left\{ \dim(S_i),dim(S_{i+1}) \right\}-1 
  \end{equation}   
  for each index $0\le i\le n-1$, where the indices are read modulo $n$. 
\end{Lemma}
\begin{Proof}
  Let $G$ be a generator matrix of $C$ and let $g_0,\dots,g_{n-1}$ denote the ordered list of columns of $G$. For each index $0\le n-1$ let $S_i$ denote the subspace 
  spanned by $g_i,g_{i+1},\dots, g_{i+b-1}$, where the indices are read modulo $n$. With this we let $\cS$ consist of the $S_i$, which have dimension at most $b$ and are contained 
  in $\PG(k-1,q)$. Consider an arbitrary non-zero codeword $c=\left(c_0,\dots,c_{n-1}\right)\in C$. There exists a unique row vector $h\in \F_q^k\backslash \{0\}$ such that $c=hG$. 
  To $h$ we can assign the set of all points $P$ such that $hP=0$, which is a hyperplane $H$ that is equal for all non-zero multiples of $h$. Note that we have 
  $\left(c_i,c_{i+1},\dots,c_{i+b-1}\right)=0$ iff $S_i\le H$, where the indices are read modulo $n$. Thus, every hyperplane $H$ in $\PG(k-1,q)$ contains at most $n-d$ elements 
  from $\cS$ and equality indeed occurs. Thus, $\cS$ is a a projective $b-(n,k,n-d)_q$ system. Inequality~(\ref{eq_dim_intersection}) directly follows from the construction since 
  $S_i\cap S_{i+1}$ is generated by $g_{i+1},g_{i+2},\dots,g_{i+b-1}$.
\end{Proof}

\begin{Example}
  \label{ex_reverse}
  In $\PG(2,2)$ consider the ordered list of subspaces
  $$ {\tiny{
    S_0=\left\langle\begin{pmatrix}1\\ 1\\ 1\end{pmatrix}\!,\begin{pmatrix}1\\ 0\\ 0\end{pmatrix}\right\rangle,
    S_1=\left\langle\begin{pmatrix}1\\ 0\\ 0\end{pmatrix}\!,\begin{pmatrix}0\\ 0\\ 1\end{pmatrix}\right\rangle,
    S_2=\left\langle\begin{pmatrix}0\\ 0\\ 1\end{pmatrix}\!,\begin{pmatrix}1\\ 0\\ 1\end{pmatrix}\right\rangle,
    S_3=\left\langle\begin{pmatrix}1\\ 0\\ 1\end{pmatrix}\!,\begin{pmatrix}1\\ 1\\ 1\end{pmatrix}\right\rangle,
    S_4=\left\langle\begin{pmatrix}1\\ 1\\ 1\end{pmatrix}\right\rangle\!}}
  $$ 
  In $\PG(2,2)$ the $[3]_2=7$ hyperplanes are indeed lines. Since the point $S_4$ is contained in the three hyperplanes $S_0$, $S_3$, 
  $\left\langle (1,1,1)^\top,(0,0,1)^\top\right\rangle$ and $S_1=S_2$ there are exactly three hyperplanes that contain two elements from 
  the multiset $\cS:=\left\{S_0,\dots,S_4\right\}$, one hyperplane that contains one element from $\cS$, and three hyperplanes that contain no element from $\cS$. 
  So, $\cS$ is a projective $2-(5,3,2)_2$ system. Noting that Inequality~(\ref{eq_dim_intersection}) is satisfied let us try to reverse Lemma~\ref{lemma_associated_projective_system} 
  and build up a generator matrix $G$ with columns $g_0,\dots,g_4$ as in the corresponding proof. Except for $S_1\cap S_2$, the intersections 
  $S_i\cap S_{i+1}$ determines a unique point, which leaves the three choices
  $$
    G_1=\begin{pmatrix} 
    11111\\
    00011\\
    00111
    \end{pmatrix}\!,
    G_2=\begin{pmatrix} 
    11111\\
    00011\\
    01111
    \end{pmatrix}\!,
    G_3=\begin{pmatrix} 
    10111\\
    00011\\
    01111
    \end{pmatrix}
  $$
  having weight enumerators $x^0y^5 +x^2y^3+2x^3y^2+2x^4y^1+2x^5y^0$, $x^0y^5 +x^2y^3+2x^3y^2+2x^4y^1+2x^5y^0$, and $x^0y^5+3x^3y^2+x^4y^1+3x^5y^0$, respectively. 
  Here, applying Lemma~\ref{lemma_associated_projective_system} to $G_i$ only gives $\cS$ for $i=3$.
  
  Now let $\cS'$ arise from $\cS$ via replacing $S_4$ by $\left\langle (1,1,1)^\top,(0,1,0)^\top\right\rangle$. If we could find a generator matrix $G'$ such that the application 
  of the construction in the proof of Lemma~\ref{lemma_associated_projective_system} gives $\cS'$, then $G'$ would generate a $[5,3,2]_2^2$ code with weight enumerator 
  $x^0y^5+x^3y^2+3x^4y^1+3x^5y^0$. However, such a generator matrix does not exist. 
\end{Example}

So, by Lemma~\ref{lemma_associated_projective_system} upper bounds for the cardinality $n$ of projective $b-(n,k,n-d)_q$ systems or the corresponding additive codes, see e.g.\ 
\cite{kurz2024additive} and the references therein, imply lower bounds for $n_q^b(k,d)$. It is interesting to note that additive codes also have to satisfy a Griesmer type inequality 
like Inequality~(\ref{eq_griesmer_bound_pair_distance}), which can always be attained with equality for sufficiently larger minimum distance $d$, see \cite{kurz2024additive} for details. 
In our situation we additionally need to ensure that the projective system $b-(n,k,n-d)_q$ system can be written as a list of subspaces satisfying Inequality~(\ref{eq_dim_intersection}), 
which is even not sufficient as shown in Exercise~\ref{ex_reverse}. For parameters $k$, $d$, $q$, and $b$ we 
call 
\begin{equation}
  \label{ie_griesmer_bound_b}
  n \ge \left\lceil \frac{g_q\!\left(k,q^{b-1}\cdot d\right)}{[b]_q} \right\rceil 
\end{equation}   
the \emph{Griesmer bound} for an $[n,k,d]_q^b$ code, c.f.\ \cite[Theorem 1]{luo2024griesmer}. The aim of the subsequent Section~\ref{sec_constructions} is to show that 
the Griesmer bound can always be attained with equality if the minimum distance $d$ is assumed to be sufficiently large.
\begin{Lemma} (C.f.\ \cite[Theorem 3]{huang2025b}) We have 
  \begin{equation}
    \left\lceil \frac{g_q\!\left(k,q^{b-1}\cdot d\right)}{[b]_q} \right\rceil
    = d+ \left\lceil \frac{g_q(k-b+1,d)-d   }{[b]_q}\right\rceil.
  \end{equation}
\end{Lemma}
\begin{Proof}
  \begin{eqnarray*}
  \!\!\!\!  && \left\lceil \frac{g_q\!\left(k,q^{b-1}\cdot d\right)}{[b]_q} \right\rceil    
    = \left\lceil \frac{\sum\limits_{i=0}^{k-1} \left\lceil\frac{d\cdot q^{b-1}}{q^i}\right\rceil}{[b]_q} \right\rceil  
    = \left\lceil\frac{ \sum\limits_{i=0}^{b-1}  \left\lceil d\cdot q^{b-i-1}\right\rceil \,+\, \sum\limits_{i=b}^{k-1} \left\lceil d\cdot q^{b-i-1}\right\rceil   }{[b]_q}\right\rceil \\ 
   \!\!\!\! &=& \left\lceil\frac{ d\cdot \sum\limits_{i=0}^{b-1}  q^i \,+\, \sum\limits_{i=1}^{k-b} \left\lceil \frac{d}{q^i}\right\rceil   }{[b]_q}\right\rceil  
    = \left\lceil\frac{ d[b]_q \,-\, d\,+\, \sum\limits_{i=0}^{k-b} \left\lceil \frac{d}{q^i}\right\rceil   }{[b]_q}\right\rceil  
    = d+ \left\lceil \frac{g_q(k-b+1,d)-d   }{[b]_q}\right\rceil 
  \end{eqnarray*}
\end{Proof}

Using a specific parameterization of the minimum distance $d$ of a linear code in the Hamming metric, the
corresponding Griesmer bound in Inequality~(\ref{eq_griesmer_bound}) can be written more
explicitly:
\begin{Lemma}
  \label{lemma_parameters_griesmer_code}
  Let $k$ and $d$ be positive integers. Write $d$ as
  \begin{equation}
    \label{eq_griesmer_representation_min_dist}
    d=\sigma q^{k-1}-\sum_{i=1}^{k-1}\varepsilon_iq^{i-1},
  \end{equation}
  where $\sigma\in\N_0$ and the $0\le\varepsilon_i<q$ are integers for all $1\le i\le k-1$. Then, Inequality~(\ref{eq_griesmer_bound})
  is satisfied with equality iff
  \begin{equation}
    \label{eq_griesmer_representation_length}
    n=\sigma[k]_q-\sum_{i=1}^{k-1}\varepsilon_i[i]_q,
  \end{equation}
  which is equivalent to
  \begin{equation}
    \label{eq_griesmer_representation_species}
    n-d=\sigma[k-1]_q-\sum_{i=1}^{k-1}\varepsilon_i[i-1]_q.
  \end{equation}
\end{Lemma}
\begin{Remark}
  Given $k$ and $d$ Equation~(\ref{eq_griesmer_representation_min_dist})
  always determines $\sigma$ and the $\varepsilon_i$ uniquely. This is
  different for Equation~(\ref{eq_griesmer_representation_species}) given
  $k$ and $n-d=s$. Here it may happen that no solution with
  $0\le \varepsilon_i\le q-1$ exists. By relaxing to
  $0\le \varepsilon_i\le q$ we can ensure existence and uniqueness is
  enforced by additionally requiring $\varepsilon_j=0$ for all $j<i$
  where $\varepsilon_i=q$ for some $i$. The same is true for
  Equation~(\ref{eq_griesmer_representation_length}) given $k$ and $n$.
  For more details we refer to \cite[Chapter 2]{govaerts2003classifications}
  which also gives pointers to Hamada's work on minihypers.
\end{Remark}

\begin{Lemma} 
  \label{lemma_griesmer_bound_period}
  $\,$\\[-4mm]
  \begin{itemize}
  \item[(a)] $g_q\!\left(k,\lambda\cdot q^{k-1}\right)=\lambda[k]_q$ for each $\lambda\in\N$.
  \item[(b)] $\left\lceil \frac{g_q\!\left(k,q^{b-1}\cdot d\right)}{[b]_q}\right\rceil = [k]_q$ for $d=q^{k-b}\cdot [b]_q$.
  \item[(c)]
    For each $\lambda,d'\in \N$ we have 
    \begin{equation}
      \left\lceil \frac{g_q\!\left(k,q^{b-1}\cdot \left(\lambda\cdot q^{k-b}\cdot [b]_q +d'\right)\right)}{[b]_q}\right\rceil
      =
      \lambda\cdot [k]_q\,+\,\left\lceil \frac{g_q\!\left(k,q^{b-1}\cdot d'\right)}{[b]_q}\right\rceil.
    \end{equation}
    \end{itemize}
\end{Lemma}
\begin{Proof}
  Parts (a), (b) directly follow from Lemma~\ref{lemma_parameters_griesmer_code}, which then imply (c).  
\end{Proof}

\section{Linear codes attaining the Griesmer bound}
\label{sec_constructions}

In this section we want to construct optimal $[n,k,d]_q^b$ codes. One possible construction is to start with a cyclic group $G$ in $\operatorname{GL}(k,q)$ generated by some element $g\in G$. 
Denoting the action of $g$ on a point $P$ by $P^g$, we can partition the set of points of $\PG(k-1,q)$ into orbits of the form $P,P^g,P^{g^2},P^{g^3},\dots,P^{g^{l-1}}$, where the length $l$ 
of the orbit can be different for different starting points $P$. We consider the sequence of points $P,P^g,P^{g^2},P^{g^3},\dots,P^{g^{l-1}}$ as a generator matrix of an $[l,k',d]_q^b$ code 
$C$ with $k'\le k$. In some cases we have $k'=k$ and $d$ is suitably large w.r.t.\ the other parameters.    

\begin{Example}
  \label{example_singer}
  In $\operatorname{GL}(5,2)$ there exist six cyclic groups of order $2^5-1=31$. Choosing a generator matrix arising from the orbit of all points with respect to a generator of the 
  cyclic group yields a $[31,5,24]_2^2$ code in all cases.
  \begin{itemize}
    \item $\left(\begin{smallmatrix}
    0 1 0 0 0 \\
    0 0 1 0 0 \\
    0 0 0 1 0 \\
    0 0 0 0 1 \\
    1 0 1 1 1
    \end{smallmatrix}\right): 
    \left(\begin{smallmatrix}
    0 1 1 0 0 1 0 0 1 1 1 1 1 0 1 1 1 0 0 0 1 0 1 0 1 1 0 1 0 0 0 \\ 
    0 0 1 1 0 0 1 0 0 1 1 1 1 1 0 1 1 1 0 0 0 1 0 1 0 1 1 0 1 0 0 \\
    0 1 1 1 1 1 0 1 1 1 0 0 0 1 0 1 0 1 1 0 1 0 0 0 0 1 1 0 0 1 0 \\
    0 1 0 1 1 0 1 0 0 0 0 1 1 0 0 1 0 0 1 1 1 1 1 0 1 1 1 0 0 0 1 \\
    1 1 0 0 1 0 0 1 1 1 1 1 0 1 1 1 0 0 0 1 0 1 0 1 1 0 1 0 0 0 0
    \end{smallmatrix}\right)\!\!,$
    \item $\left(\begin{smallmatrix}
    0 1 0 0 0 \\
    0 0 1 0 0 \\
    0 0 0 1 0 \\
    0 0 0 0 1 \\
    1 1 1 1 0
    \end{smallmatrix}\right):
    \left(\begin{smallmatrix}
    0 1 0 1 1 0 1 0 1 0 0 0 1 1 1 0 1 1 1 1 1 0 0 1 0 0 1 1 0 0 0 \\
    0 1 1 1 0 1 1 1 1 1 0 0 1 0 0 1 1 0 0 0 0 1 0 1 1 0 1 0 1 0 0 \\
    0 1 1 0 0 0 0 1 0 1 1 0 1 0 1 0 0 0 1 1 1 0 1 1 1 1 1 0 0 1 0 \\
    0 1 1 0 1 0 1 0 0 0 1 1 1 0 1 1 1 1 1 0 0 1 0 0 1 1 0 0 0 0 1 \\
    1 0 1 1 0 1 0 1 0 0 0 1 1 1 0 1 1 1 1 1 0 0 1 0 0 1 1 0 0 0 0 
    \end{smallmatrix}\right)\!\!,$
    \item $\left(\begin{smallmatrix}
    0 1 0 0 0 \\
    0 0 1 0 0 \\
    0 0 0 1 0 \\
    0 0 0 0 1 \\
    1 1 1 0 1    
    \end{smallmatrix}\right):
    \left(\begin{smallmatrix}
    0 1 1 1 0 0 1 1 0 1 1 1 1 1 0 1 0 0 0 1 0 0 1 0 1 0 1 1 0 0 0 \\
    0 1 0 0 1 0 1 0 1 1 0 0 0 0 1 1 1 0 0 1 1 0 1 1 1 1 1 0 1 0 0 \\
    0 1 0 1 0 1 1 0 0 0 0 1 1 1 0 0 1 1 0 1 1 1 1 1 0 1 0 0 0 1 0 \\
    0 0 1 0 1 0 1 1 0 0 0 0 1 1 1 0 0 1 1 0 1 1 1 1 1 0 1 0 0 0 1 \\
    1 1 1 0 0 1 1 0 1 1 1 1 1 0 1 0 0 0 1 0 0 1 0 1 0 1 1 0 0 0 0
    \end{smallmatrix}\right)\!\!,$
    \item $\left(\begin{smallmatrix}
    0 1 0 0 0 \\
    0 0 1 0 0 \\
    0 0 0 1 0 \\
    0 0 0 0 1 \\
    1 1 0 1 1
    \end{smallmatrix}\right):     
    \left(\begin{smallmatrix}
    0 1 1 0 1 0 1 0 0 1 0 0 0 1 0 1 1 1 1 1 0 1 1 0 0 1 1 1 0 0 0 \\
    0 1 0 1 1 1 1 1 0 1 1 0 0 1 1 1 0 0 0 0 1 1 0 1 0 1 0 0 1 0 0 \\
    0 0 1 0 1 1 1 1 1 0 1 1 0 0 1 1 1 0 0 0 0 1 1 0 1 0 1 0 0 1 0 \\
    0 1 1 1 1 1 0 1 1 0 0 1 1 1 0 0 0 0 1 1 0 1 0 1 0 0 1 0 0 0 1 \\
    1 1 0 1 0 1 0 0 1 0 0 0 1 0 1 1 1 1 1 0 1 1 0 0 1 1 1 0 0 0 0
    \end{smallmatrix}\right)\!\!,$
    \item $\left(\begin{smallmatrix}
    0 1 0 0 0 \\
    0 0 1 0 0 \\
    0 0 0 1 0 \\
    0 0 0 0 1 \\
    1 0 0 1 0
    \end{smallmatrix}\right):
    \left(\begin{smallmatrix}
    0 1 0 1 0 1 1 1 0 1 1 0 0 0 1 1 1 1 1 0 0 1 1 0 1 0 0 1 0 0 0 \\
    0 0 1 0 1 0 1 1 1 0 1 1 0 0 0 1 1 1 1 1 0 0 1 1 0 1 0 0 1 0 0 \\
    0 0 0 1 0 1 0 1 1 1 0 1 1 0 0 0 1 1 1 1 1 0 0 1 1 0 1 0 0 1 0 \\
    0 1 0 1 1 1 0 1 1 0 0 0 1 1 1 1 1 0 0 1 1 0 1 0 0 1 0 0 0 0 1 \\
    1 0 1 0 1 1 1 0 1 1 0 0 0 1 1 1 1 1 0 0 1 1 0 1 0 0 1 0 0 0 0 
    \end{smallmatrix}\right)\!\!,$
    \item $\left(\begin{smallmatrix}
    0 1 0 0 0 \\
    0 0 1 0 0 \\
    0 0 0 1 0 \\
    0 0 0 0 1 \\
    1 0 1 0 0
    \end{smallmatrix}\right):
    \left(\begin{smallmatrix}
    0 1 0 0 1 0 1 1 0 0 1 1 1 1 1 0 0 0 1 1 0 1 1 1 0 1 0 1 0 0 0 \\
    0 0 1 0 0 1 0 1 1 0 0 1 1 1 1 1 0 0 0 1 1 0 1 1 1 0 1 0 1 0 0 \\
    0 1 0 1 1 0 0 1 1 1 1 1 0 0 0 1 1 0 1 1 1 0 1 0 1 0 0 0 0 1 0 \\
    0 0 1 0 1 1 0 0 1 1 1 1 1 0 0 0 1 1 0 1 1 1 0 1 0 1 0 0 0 0 1 \\
    1 0 0 1 0 1 1 0 0 1 1 1 1 1 0 0 0 1 1 0 1 1 1 0 1 0 1 0 0 0 0 
    \end{smallmatrix}\right)\!\!.$
  \end{itemize}
  Interestingly enough, deleting one arbitrary column from the generator matrices yields an $[30,5,22]_2^2$ code in all cases, i.e.\ the minimum symbol pair distance is decreased by two 
  while the length is just increased by one.
\end{Example}

In $\operatorname{GL}(k,q)$ the maximum order of a cyclic group is $[k]_q$ and it acts transitively on the set of points in $\PG(k-1,q)$. Those groups exist for all parameters and are 
called \emph{Singer groups}, see e.g.~\cite{hestenes1970singer}.

\begin{Proposition}
  \label{prop_singer_construction}
  For $k\ge b\ge 2$ there exists an $[n,k,d]_q^b$ code $C$ with $n=[k]_q$ and $d=[b]_q\cdot q^{k-b}$.
\end{Proposition}
\begin{Proof}
  Let $M\in \operatorname{GL}(k,q)$ be a $k\times k$ matrix over $\F_q$ that generates a cyclic group of order $n=[k]_q$, i.e.\ a \emph{Singer cycle}. Denoting the 
  first unit vector by $e_1$ we construct a generator matrix $G$ for $C$ by choosing the sequence of columns $M^0 e_1,M^1 e_1,\dots,M^{n-1}e_1$. Note that $M$ acts transitively on the
  set of $n$ points of $\PG(k-1,q)$. Set 
  \begin{equation}
    S_i:=\left\langle M^i e_1,M^{i+1}e_1,\dots, M^{i+b-1}e_1\right\rangle
  \end{equation} 
  for all $0\le i\le n-1$, where the indices are read modulo $n$. Let $P:=M^i e_1$ and suppose that there exist $\lambda_0,\dots,\lambda_{b-2}\in\F_q$ such that $M^{b-1}P=\sum_{i=0}^{b-2} M^i P$.  
  Then, $M^0P, M^1P, \dots $ is contained in $\left\langle M^0P,M^1P,\dots,M^{b-2}P\right\rangle$, which is impossible since $M$ acts transitively on the set of points and $k\ge b$. Thus, 
  we have $\dim(S_i)=b$ for all $0\le i\le n-1$. 
  
  Let $\cM:=\sum_{i=0}^{n-1} \chi_{S_i}$, so that $\# \cM=n[b]_q=[k]_q[b]_q$. Note that each point $P$ in $S_i$ can be written as 
  $\left\langle \sum_{j=0}^{b-1} \lambda_jM^{i+j}e_1\right\rangle=M^i \cdot \left\langle \sum_{j=0}^{b-1} \lambda_jM^{j}e_1\right\rangle$, where 
  $\left(\lambda_0,\dots,\lambda_{b-1}\right)\in\F_q^b\backslash\{0\}$ is uniquely determined. Thus, the transitivity of $M$ on the set of points implies $\cM(P)=\cM(P')$
  for all pairs of points $P,P'$. Counting points then yields $\cM(P)=[b]_q$ for every point $P$. Since each of the $b$-spaces $S_i$ intersects a given hyperplane $H$ in 
  either $[b]_q$ or $[b-1]_q$ points and $[b]_q=q^{b-1}+[b-1]_q$ we have
  \begin{equation}    
    [k-1]_q[b]_q=\cM(H) = [k]_q[b-1]_q + s\cdot q^{b-1},
  \end{equation}  
  where $s$ denotes the number of indices $0\le i\le n-1$ with $S_i\le H$. Since
  \begin{eqnarray*}
    [k-1]_q[b]_q - [k]_q[b-1]_q &=& \frac{\left(q^{k-1}-1\right)\cdot\left(q^b-1\right)-\left(q^{k}-1\right)\cdot\left(q^{b-1}-1\right)}{(q-1)^2} \\ 
    &=& \frac{q^k + q^{b-1} -q^{k-1}- q^b}{(q-1)^2}=\frac{q^{k-1}-q^{b-1}}{q-1}=q^{b-1}\cdot [k-b]_q, 
  \end{eqnarray*}
  we conclude $s=[k-b]_q$. Since
  \begin{eqnarray*}
    n-s &=& [k]_q - [k-b]_q =\frac{q^k-q^{k-b}}{q-1} =q^{k-b}\cdot \frac{q^b-1}{q-1}= q^{k-b}\cdot [b]_q,
  \end{eqnarray*} 
  the constructed code has minimum distance $\db(C)=[b]_q\cdot q^{k-b}$.   
\end{Proof}
In the Hamming metric the constructed code $C$ is just a simplex code. The minimum distance of those codes w.r.t.\ the $b$-symbol metric 
was studied in \cite{ma2021symbol} for the special case $b\le q-1$. Due to Lemma~\ref{lemma_griesmer_bound_period}.(b) the codes from 
Proposition~\ref{prop_singer_construction} attain the Griesmer bound of Inequality~(\ref{ie_griesmer_bound_b}). Appending $\lambda$ copies 
of the corresponding generator matrix yields $[n,k,d]_q^b$ codes with $n=\lambda\cdot [k]_q$ and $d=\lambda\cdot [b]_q\cdot q^{k-b}$, so that 
Lemma~\ref{lemma_griesmer_bound_period}.(c) gives 
\begin{equation}
  \lim_{d\to\infty} n_q^b(k,d) / \left\lceil \frac{g_q\!\left(k,q^{b-1}\cdot d\right)}{[b]_q}\right\rceil = 1.
\end{equation}

\begin{Lemma}
  \label{lemma_sum_special}
  Let $G_1$ be a generator matrix of an $\left[n_1,k,d_1\right]_q^b$ code and $G_2$ be a generator matrix of an $\left[n_2,k,d_2\right]_q^b$ code. If the first 
  $b-1$ columns of $G_1$ and $G_2$ coincide, then appending $G_2$ to $G_1$ is the generator matrix of an $\left[n_1+n_2,k,d_1+d_2\right]_q^b$ code.  
\end{Lemma}

\begin{Definition}
  We call an $[n,k]_q^b$ code $C$ \emph{faithful} if the restriction of the codewords to the coordinates $i,i+1,\dots i+b-1$ (read modulo $n$) form an
  $[b,b]_q^b$ code for each index $0\le i\le n-1$.
\end{Definition}

In other words, an $[n,k,d]_q^b$ code is faithful iff the projective $b-(n,k,n-d)_q$ system is faithful, where the generator matrix can be chosen arbitrarily.
If $\cS$ is an arbitrary $h-(n,k,s)_q$ system, then we can obtain a faithful $h-(n,k,\le s)_q$ system $\cS'$ by replacing each element $S\in\cS$ with $\dim(S)<h$
by a $b$-space $S'$ with $S\le S'$. A similar results also holds for $[n,k]_q^b$ codes.

\begin{Lemma}
  \label{lemma_faitful}
  Let $C$ be an $[n,k,d]_q^b$ code with $k\ge b$. Then, there exists a faithful $[n,k,\ge d]_q^b$ code $C'$.
\end{Lemma} 
\begin{Proof}
  We iteratively construct a generator matrix $G_i$ of an $[n,k,\ge d]_q^b$ code such that $i$ subsequent columns span an $i$-space for all $0\le i\le b$, where the column indices 
  are read modulo $n$. Going from $G_i$ to $G_{i+1}$ we modify just one column, so that the length remain $n$ during our entire construction. The improvement property we maintain 
  during the construction is the following. On the way from $G_i$ to $G_{i+1}$ we assume that we have already constructed a generator matrix $U$ of an an $[n,k,\ge d]_q^b$ code such 
  that $i$ subsequent columns span an $i$-space. In the next step we construct a generator matrix $U'$ of an an $[n,k,\ge d]_q^b$ code such that $i$ subsequent columns span an $i$-space, 
  $\left\langle u_j,u_{j+1},\dots,u_{j+i}\right\rangle\le \left\langle u_j',u_{j+1}',\dots,u_{j+i}'\right\rangle$ for all $0\le j\le n-1$ and that the dimension increases at least once, 
  where $u_0,\dots, u_{n-1}$ are the columns of $U$, $u_0',\dots,u_{n-1}'$ are the columns of $U'$, and the indices are read modulo $n$. Let us denote the index where we change a column 
  by $h$, i.e., we have $u_j=u_j'$ for all $0\le j\le n-1$ with $j\neq h$. Let us first observe that $U'$ generates an $[n,k]_q^b$ code if $U$ does. To this end we consider 
  \begin{eqnarray*}
    && \langle u_0',u_1',\dots,u_{n-1}'\rangle \\ 
    &=& \big\langle \langle u_h',u_{h+1}',\dots u_{h+i}',\rangle, \langle u_j' | 0\le j\le n-1, j\notin [h,h+i], j\notin [h-n,h+i-n] \rangle\big \rangle \\
    &=& \big\langle \langle u_h',u_{h+1}',\dots u_{h+i}',\rangle, \langle u_j | 0\le j\le n-1, j\notin [h,h+i], j\notin [h-n,h+i-n] \rangle\big \rangle\\ 
    &\ge & \big\langle \langle u_h,u_{h+1},\dots u_{h+i},\rangle, \langle u_j | 0\le j\le n-1, j\notin [h,h+i], j\notin [h-n,h+i-n] \rangle\big \rangle \\
    &=& \langle u_0,u_1,\dots,u_{n-1}\rangle.
  \end{eqnarray*} 
  We can perform the same computation for the span of $b$ subsequent columns of $U$ and $U'$ to conclude that the minimum $b$-symbol distance does not decrease. 
  
  \medskip
  
  As $G_0$ we can take an arbitrary generator matrix $G$ of $C$. For the construction of $G_1$ we iteratively replace each occurring zero vector by an arbitrary non-zero vector.
  Clearly, we have $\left\langle u_j\right\rangle\le \left\langle u_j'\right\rangle$ for all $0\le j\le n-1$ since $u_j=u_j'$ if $j\neq h$ and $u_h'$ is the zero vector. 
  Moreover, we have $\dim(\langle u_h'\rangle)=1>0=dim(\langle u_h\rangle)$. Here we assume that we directly set $G_{i+1}=G_i$, if $G_i$ already has the desired property of $G_{i+1}$, 
  and do not consider modifications from $U$ to $U'$. 
  
  Now we assume that we have a generator matrix $U$ such that all $i$ subsequent columns span an $i$-space and that there exists a index $0\le j\le n-1$ such that 
  the columns $u_j,\dots, u_{j+i}$ also span an $i$-space and indeed not span an $(i+1)$-span. So, we have $\langle u_j,\dots,u_{j+i-1}\rangle =\langle u_{j+1},\dots,u_{j+i}\rangle$.
  Since the span of all $n$ columns has dimension $k$, we can assume $\langle u_{j+1},\dots,u_{j+i}\rangle \neq\langle u_{j+2},\dots,u_{j+i+1}\rangle$ for $i<b\le k$ by possibly increasing 
  the initial index $j$. With this we choose $h=j+i$ (modulo $n$) and set $u'_h=u_h+u_{h+1}$. For brevity we set $\cS_l:=\langle u_l,\dots,u_{l+i-1}\rangle$, 
  $\cS_l':=\langle u_l',\dots,u_{l+i-1}'\rangle$, $\cT_l:=\langle u_l,\dots,u_{l+i}\rangle$, and $\cT_l':=\langle u_l',\dots,u_{l+i}'\rangle$ for all $0\le l\le n-1$, so that 
  e.g.\ $\cS_j=\cS_{j+1}\neq \cS_{j+2}$ and $\dim(\cT_j)=i$. Since $\langle u_h,u_{h+1}\rangle=\langle u_h',u_{h+1}'\rangle$ we have $\cS_l=\cS_l'$ for $l\notin \{j+1,j+1-n\}$ 
  and $\cT_l=\cT_l'$ $l\notin \{j,j-n\}$. Since $\cS_{j+1}\neq \cS_{j+2}$ we have $\cT_j'>\cT_j$, $\dim(\cS_{j+1}')=i$, $\cS_j'\neq \cS_{j+1}'$, and $\cS_{j+1}'\neq \cS_{j+2}'$. 
  Thus, $U'$ satisfies the required property.   
\end{Proof}

We remark that the code constructed in Proposition~\ref{prop_singer_construction} is faithful.

\begin{Lemma}
  \label{lemma_addition}
  $$
    n_q^b\!\left(k,d_1+d_2\right )\le n_q^b\!\left(k,d_1\right)+n_q^b\!\left(k,d_2\right) 
  $$
\end{Lemma}
\begin{Proof}
  Let $C_i$ be faithful $\left[n_i,k,d_i\right]_q^b$ codes with $n_i=n_q^b\!\left(k,d_i\right)$ and $G_i$ be corresponding generator matrices for $i\in\{1,2\}$, see Lemma~\ref{lemma_faitful}. 
  Multiplying $G_i$ by a suitable matrix in $\operatorname{GL}(k,q)$ we obtain a generator matrix $G_i'$ of a faithful $\left[n_i,k,d_i\right]_q^b$ code starting with a $k\times k$ unit matrix 
  $I_k$, where $i\in \{1,2\}$. Applying Lemma~\ref{lemma_sum_special} to $G_1'$ and $G_2'$ yields a generator matrix of an  $\left[n_1+n_2,k,d_1+d_2\right]_q^b$ code.    
\end{Proof}

For the $2$-symbol metric we can directly state the resulting minimum distance for the concatenation of two arbitrary generator matrices. For simplicity we state the observation in terms of codewords.  
\begin{Lemma}
  \label{lemma_concatenation}
  Let $a\in \F_q^{n_1}$ and $b\in \F_q^{n_2}$, where $n_1,n_2\ge 2$, such that $\operatorname{wt}_2(a)=d_1$ and $\operatorname{wt}_2(b)=d_2$. Construct $c\in\F_q^{n_1+n_2}$ as the concatenation of 
  $a$ and $b$ -- written $c=a|b$.   Writing $\star$ for an arbitrary element in $\F_q\backslash\{0\}$ we have the following four cases for $a$:
  \begin{itemize}
    \item[(a)] $a=\left(0,a_1,\dots,a_{n-2},0\right)$;
    \item[(b)] $a=\left(0,a_1,\dots,a_{n-2},\star\right)$;
    \item[(c)] $a=\left(\star,a_1,\dots,a_{n-2},0\right)$;
    \item[(d)] $a=\left(\star,a_1,\dots,a_{n-2},\star\right)$.
  \end{itemize}  
  Similar for $b$:
  \begin{itemize}
    \item[(i)] $b=\left(0,b_1,\dots,b_{n-2},0\right)$;
    \item[(ii)] $b=\left(0,b_1,\dots,b_{n-2},\star\right)$;
    \item[(iii)] $b=\left(\star,b_1,\dots,b_{n-2},0\right)$;
    \item[(iv)] $b =\left(\star,b_1,\dots,b_{n-2},\star\right)$.
  \end{itemize}
  With this, we have $\operatorname{wt}_2(c)=\operatorname{wt}_2(a)+\operatorname{wt}_2(b)-1$ if we are in case (b).(iii) or case (c).(ii) and 
  $\operatorname{wt}_2(c)=\operatorname{wt}_2(a)+\operatorname{wt}_2(b)$ in all other cases. 
\end{Lemma}

\begin{Corollary}
  \label{corollary_concatenation}
  Let $n\ge 2$, $k\ge 2$, and $t\ge 1$ be integers. For each $c\in\F_q^n$ with $n\ge 2$ we have $\operatorname{wt}_2(\overset{t\text{ times}}{\overbrace{c|\cdots|c}})=t\cdot\operatorname{wt}_2(c)$. 
  For each full rank matrix $G\in \F_q^{k\times n}$ we have $\operatorname{d}_2\!\big(\!\operatorname{rowspan}\!\big( \overset{t\text{ times}}{\overbrace{G|\dots|G}}\big)\big)=t\cdot 
  \operatorname{d}_2\!\big(\!\operatorname{rowspan}\!\big(G\big)\big)$. 
\end{Corollary}

\bigskip

Our next aim is to show
\begin{equation}
  \lim_{d\to\infty} n_q^b(k,d) -\left\lceil \frac{g_q\!\left(k,q^{b-1}\cdot d\right)}{[b]_q}\right\rceil = 0.
\end{equation}
\begin{Lemma}
  Let $G_i$ be generator matrices of faithful $\left[n_i,k,d_i\right]_q^b$ codes for $0\le i<l$ such that the last $b-1$ columns of $G_i$ coincide with the first $b-1$ columns of
  $G_{i+1}$ for all $0\le i<l$, where the indices are read modulo $l$. Then, the concatenation of the matrices $G_0,G_1,\dots, G_{l-1}$ is the generator matrix of a
  faithful $\left[\sum_{i=0}^{l-1} n_i,k,\sum_{i=0}^{l-1} d_i\right]_q^b$ code.
\end{Lemma} 
    
\begin{Example}
  \label{example_chain_motivation}
  The matrix
  $$
    M=
    \begin{pmatrix}
    0 & 1 & 0 & 0 & 0 \\
    0 & 0 & 1 & 0 & 0 \\
    0 & 0 & 0 & 1 & 0 \\
    0 & 0 & 0 & 0 & 1 \\
    1 & 0 & 1 & 1 & 1
    \end{pmatrix}\in\operatorname{GL}(5,2)
  $$
  generates the orbit of points
  $$
  G=\begin{pmatrix}
  1000011001001111101110001010110\\
  0000110010011111011100010101101\\
  0001100100111110111000101011010\\
  0011001001111101110001010110100\\
  0110010011111011100010101101000 
  \end{pmatrix}
  $$
  when starting from the first unit vector. We write $v_j$ for the vector $v_j\in\F_2^5\backslash\{0\}$ which coincides with the binary expansion of $j\le i\le 31$. So for the subspaces 
  $\cS_i$ according to Lemma~\ref{lemma_associated_projective_system} we e.g.\ have $\cS_0=\langle v_{16},v_1,v_3\rangle$, $\cS_{29}=\langle v_{20}, v_{8},v_{16}\rangle$, and 
  $\cS_{30}=\langle v_ 8 ,v_{16},v_1\rangle$. More precisely, the first two columns of $G$ equal $v_{16}$, $v_1$ and the last column of $G$ equals $v_8$. 
  Setting $\cS_{29}'=\langle v_{20}, v_{8},v_{16}+v_{20}\rangle$, $\cS_{30}'=\langle v_ 8 ,v_{16}+v_{20},v_1\rangle$, $\cS_{29}''=\langle v_{20}, v_{8},v_{16}\rangle$, and 
  $\cS_{30}''=\langle v_ 8 ,v_{16},v_1+v_{20}\rangle$, we note $\cS_{29}=\cS_{29}'=\cS_{29}''$ and $\cS_{30}=\cS_{30}'=\cS_{30}''$. So, we may either append the vectors $v_{16},v_1$ to 
  $G$ or the vectors $v_{16}+v_{20},v_1$ or the vectors $v_{16},v_1+v_{20}$ so that the first $31$ planes coincide.    
\end{Example}    
\begin{Definition}
  A \emph{$b$-chain of length $n$ over $\F_q^k$} is a list of vectors $v_0,\dots,v_{n+b-2}\in\F_q^k$. The sublist $v_0,v_1,\dots,v_{b-2}$ is called the \emph{start} and the 
  sublist $v_n,v_{n+1},\dots,v_{n+b-2}$ is called the \emph{end}. The \emph{associated projective $b-(n,k,s)_q$ system} $\cM$ consists of the spaces $\left\langle v_i,\dots v_{i+b-1}\right\rangle$ 
  for $0\le i\le n-1$ and $s$ is the maximum number of those subspaces that are contained in some hyperplane of $\PG(k-1,q)$. Two $b$-chains of length $n$ over $\F_q^k$  are called 
  \emph{equivalent} if their associated projective $b-(n,k,s)_q$ systems coincide.  
\end{Definition}

Directly from the definition we verify:
\begin{Lemma}
  \label{lemma_chain_concatenation}
  Let $v_0,\dots,v_{n+b-2}\in\F_q^k$ be a $b$-chain of length $n$ with associated projective $n-(n,k,s)_q$ system $\cM$ and $v_0',\dots,v_{n'+b-2}'\in\F_q^k$ be a $b$-chain of length $n'$ 
  with associated projective $n-(n',k,s')_q$ system $\cM'$. If $v_n,\dots v_{n+b-2}=v_0',\dots,v_{b-2}'$, i.e.\ if the end of the first chain equals the end of the second chain, then 
  $v_0,\dots,v_{n-1},v_0',\dots,v_{n'+b-2}$ is a $b$-chain of length $n+n'$ with associated projective $b-(n+n',k,s+s')_q$ system $\cM+\cM'$.    
\end{Lemma}

\begin{Lemma}
  \label{lemma_simple_chain}
  Let $v_0,\dots,v_{b-2}\in \F_q^k$ span a $(b-1)$-space, $h\in\{0,\dots,b-2\}$, $v_0',\dots,v_{b-2}'\in \F_q^k$ with $v_i'=v_i$ for $i\neq h$, $\dim(\langle v_0',\dots,v_{b-2}'\rangle)=b-1$, and 
  $\langle v_0,\dots,v_{b-2}\rangle\neq \langle v_0',\dots,v_{b-2}\rangle$. Then, there exist a $b$-chain of length $[k]_q$ with start $v_0,\dots,v_{b-2}$, end $v_0',\dots,v_{b-2}'$, and 
  associated projective $b-\left([k]_q,k,[k-b]_q\right)_q$ system.    
\end{Lemma}
\begin{Proof}
  First we note $\dim(\langle v_0,\dots,v_{b-2},v_h'\rangle)=b$. From Proposition~\ref{prop_singer_construction} we conclude the existence of a faithful $\left([k]_q,k,[b]_q\cdot q^{k-b}\right)_q^b$ 
  code $C$. Let $G$ denote a generator matrix of $G$ and $g_0,\dots, g_{n-1}$ be its columns, where $n:=[k]_q$. Noting that $\dim(\langle g_0,\dots,g_{b-2},g_{n-1}\rangle)=b$ and 
  $\dim(\langle v_0,\dots,v_{b-2},v_h+v_h'\rangle)=b$ we conclude the existence of a matrix $M\in\operatorname{GL}(k,q)$ that maps $g_i$ to $v_i$ for $0\le i\le b-2$ and $g_{n-1}$ to $v_h+v_h'$. 
  With this, $G'=M\cdot G$ starts with $v_0,\dots,v_{b-2}$, ends with $v_h+v_h'$, and generates a faithful $\left([k]_q,k,[b]_q\cdot q^{k-b}\right)_q^b$ code. From this we obtain 
  a $b$-chain with length $n$ over $\F_q^k$ with start $v_0,\dots, v_{b-2}$, end $v_0,\dots, v_{b-2}$, and associated projective $b-(n,k,s)_q$ system, where $s=[k]_q-[b]_q\cdot q^{k-b}=[k-b]_q$. Since 
  $v_{n-1}=v_h+v_h'$ this chain is equivalent to a $b$-chain with length $n$ over $\F_q^k$ with start $v_0,\dots, v_{b-2}$, end $v_0',\dots, v_{b-2}'$, and associated projective $b-(n,k,s)_q$ 
  system, where $s=[k-b]_q$.    
\end{Proof}

\begin{Lemma}
  \label{lemma_chain}
  Let $v_0,\dots,v_{b-2}\in \F_q^k$ and $v_0',\dots,v_{b-2}'\in \F_q^k$ both span a $(b-1)$-space. Then, there exists a $b$-chain of length $\lambda\cdot [k]_q$ with start $v_0,\dots,v_{b-2}$, end 
  $v_0',\dots,v_{b-2}'$, and associated projective $b-\left(\lambda\cdot [k]_q,k,\lambda\cdot[k-b]_q\right)_q$ system for some positive integer $\lambda$.  
\end{Lemma}
\begin{Proof}
  If $v_0'\notin \left\langle v_0,\dots,v_{b-2}\right\rangle$ (or $v_0=v_0'$), then Lemma~\ref{lemma_simple_chain} yields a $b$-chain with length $[k]_q$ over $\F_q^k$ that has start 
  $v_0,\dots, v_{b-2}$, end $v_0',v_1,\dots,v_{b-2}$, and an associated projective $b-\left([k]_q,k,[k-b]_q\right)_q$ system. Now assume 
  $v_0'\in \left\langle v_0,\dots,v_{b-2}\right\rangle$ and choose an index $0\le i\le b-2$ such that $v_0'\notin \left\langle v_0,\dots,v_{i-1},v_{i+1},\dots,v_{b-2}\right\rangle$ as 
  well as a vector $x\in\F_q\backslash\{0\}$ that is not contained in $\left\langle v_0,\dots,v_{i-1},v_{i+1},\dots,v_{b-2},v_0'\right\rangle$. By Lemma~\ref{lemma_simple_chain} there 
  exists a $b$-chain $C_1$ with length $[k]_q$ over $\F_q^k$ that has start $v_0,\dots, v_{b-2}$, end $v_0,\dots,v_{i-1},x,v_{i+1},\dots,v_{b-2}$, and an associated projective 
  $b-\left([k]_q,k,[k-b]_q\right)_q$ system. Now let $C_2$ be a $b$-chain of length $[k]_q$ over $\F_q^k$ with start $v_0,\dots,v_{i-1},x,v_{i+1},\dots,v_{b-2}$, 
  end $v_0',v_1,\dots,v_{i-1},x,v_{i+1},\dots,v_{b-2}$, and an associated projective $b-\left([k]_q,k,[k-b]_q\right)_q$ system. Applying Lemma~\ref{lemma_chain_concatenation} 
  to the chains $C_1$ and $C_2$ gives a $b$-chain with length $2\cdot [k]_q$ over $\F_q^k$ that has start $v_0,\dots, v_{b-2}$, end $v_0',v_1\dots,v_{i-1},x,v_{i+1},\dots,v_{b-2}$, and an 
  associated projective $b-\left(2\cdot [k]_q,k,2\cdot[k-b]_q\right)_q$ system.
  
  Suppose we have already constructed a $b$-chain $C_1$ with length $\lambda'\cdot [k]_q$ over $\F_q^k$ that has start $v_0,\dots, v_{b-2}$, end $v_0',\dots,v_i',u_{i+1},\dots,u_{b-2}$, 
  where $u_{i+1}\neq v_{i+1}'$, and an associated projective $b-\left(\lambda'\cdot [k]_q,k,\lambda'\cdot[k-b]_q\right)_q$ system for some positive integer $\lambda'$. 
  If $v_i'\notin \left\langle v_0',\dots,v_i',u_  {i+1},\dots, u_{b-2}\right\rangle$, then Lemma~\ref{lemma_simple_chain} yields a $b$-chain $C_2$ with length $[k]_q$ over $\F_q^k$ that has start 
  $v_0',\dots,v_i',u_{i+1},\dots,u_{b-2}$, end $v_0',\dots,v_{i+1}',u_{i+2},\dots,u_{b-2}$, and an associated projective $b-\left([k]_q,k,[k-b]_q\right)_q$ system. 
  Applying Lemma~\ref{lemma_chain_concatenation} to the chains $C_1$ and $C_2$ gives a $b$-chain with length $(\lambda'+1)\cdot[k]_q$ and the same properties as the initial chain while the value of $i$ 
  is increased. Now assume $v_{i+1}'\in \left\langle v_0',\dots,v_{i}',u_{i+1},\dots,u_{b-2}\right\rangle$ and choose an index $0\le j\le b-2$ such that $v_{i+1}'\notin 
  \left\langle v_0',\dots,v_{i},u_{i+1},\dots,u_{j-1},u_{j+1},\dots,u_{b-2}\right\rangle$ as well as a vector $x\in\F_q\backslash\{0\}$ that is not contained in 
  $\left\langle v_0',\dots,v_{i+1}',u_{i+1},\dots,u_{j-1},u_{j+1},\dots,u_{b-2}\right\rangle$. Using Lemma~\ref{lemma_simple_chain} we can construct  
  a $b$-chain $C_2'$ with length $[k]_q$ over $\F_q^k$ that has start $v_0',\dots,v_i',u_{i+1},\dots, u_{b-2}$, end $v_0',\dots,v_i',u_{i+1},\dots,u_{j-1},x,u_{j+1},\dots,u_{b-2}$, 
  and an associated projective $b-\left([k]_q,k,[k-b]_q\right)_q$ system. Using Lemma~\ref{lemma_simple_chain} we construct a $b$-chain $C_3'$ with length $[k]_q$ over $\F_q^k$ that has start 
  $v_0',\dots,v_i',u_{i+1},\dots,u_{j-1},x,u_{j+1},\dots,u_{b-2}$, end $v_0',\dots,v_{i+1}',u_{i+2},\dots,u_{j-1},x,u_{j+1},\dots,u_{b-2}$, and associated proj.\ 
  $b-\left([k]_q,k,[k-b]_q\right)_q$ system. Applying Lemma~\ref{lemma_chain_concatenation} to the chains $C_1$ and $C_2'$ and then applying Lemma~\ref{lemma_chain_concatenation} 
  another time to the previous result and $C_3'$ gives a $b$-chain with length $(\lambda'+2)\cdot [k]_q$ and the same properties as the initial chain while the value of $i$ 
  is increased. After an iterative application of this construction we end up with $i=b-2$ and have obtained the final desired chain.  
\end{Proof}

\begin{Example}
  We continue Example~\ref{example_chain_motivation}. Using the introduced notation and $v_{20}+v_{16}=v_4$ we start from a $3$-chain $C_1$ of length $31$ with start $v_{16},v_1$, 
  end $v_4,v_1$, and associated projective $3-(31,5,3)_2$ system $\cM_1$. From Lemma~\ref{lemma_simple_chain} we can obtain a $3$-chain $C_2$ with length $31$ having start $v_4,v_1$,
  end $v_4,v_{16}$, and associated projective $3-(31,5,3)_2$ system $\cM_2$. Applying Lemma~\ref{lemma_chain_concatenation} to $C_1$ and $C_2$ gives a $3$-chain $C$ of length $62$ 
  with start $v_{16},v_1$, end $v_4,v_{16}$, and associated projective $3-(62,5,6)_2$ system $\cM$. Note that Lemma~\ref{lemma_chain} guarantees the existence of a $3$-chain of length 
  $31\lambda$ with start $v_{16},v_1$, end $v_4,v_{16}$, and associated projective $3-(31\lambda,5,3\lambda)_2$ system for some positive integer $\lambda$. Now let 
  $S=\left\langle v_4,v_{16},v_1\right\rangle$ be a $3$-space. Reordering the stated basis of $S$ we can interprete $S$ as a $3$-chain of length $1$ with start $v_4,v_{16}$, 
  $v_{16},v_1$,, and associated $3-(1,5,1)_2$ system. Appending these two chains yields a $3$-chain with length $63$ over $\F_2^5$ with equal start and end, so that it can be 
  interpreted as an $[63,5]_2^3$ code.     
\end{Example}

Now we are ready to prove that the Griesmer bound can be attained for $[n,k,d]_q^b$ codes if the minimum distance $d$ is sufficiently large. The main idea is to start 
with a matching additive code that attains the Griesmer bound and to consider its corresponding projective system of $b$-spaces. Those $b$-spaces then are linked to together 
with suitable chains, constructed in Lemma~\ref{lemma_chain}, to a large chain having the same start and end that can then be interpreted as $[n,k,d]_q^b$ code. On the technical 
side we have to deal with the analysis of the minimum distance and the periodicity pattern of the Griesmer bound, see Lemma~\ref{lemma_griesmer_bound_period}(c).  

\begin{Theorem}
  \label{thm_main}
  Given parameters $k$, $q$, and $b$ we have 
  \begin{equation}
    n_q^b(k,d)
    =
    \left\lceil \frac{g_q\!\left(k,q^{b-1}\cdot d\right)}{[b]_q}\right\rceil
  \end{equation} 
  for all sufficiently large $d$. 
\end{Theorem}
\begin{Proof}
  Consider $1\le d'\le q^{k-b}\cdot [b]_q$ separately. From \cite[Theorem 4]{kurz2024additive} we conclude he existence of a constant $\lambda\in\N$ such that
  there exists a faithful projective $h-(n_\lambda,k,n_\lambda-d_\lambda)_q$ system $\cS_\lambda$ with $d_\lambda=\lambda\cdot q^{k-b}\cdot [b]_q +d'$ 
  and $n_\lambda= \left\lceil \frac{g_q\!\left(k,q^{b-1}\cdot d_\lambda\right)}{[b]_q}\right\rceil\!$. Now we interprete the $n_\lambda$ $b$-spaces of $\cS_\lambda$ 
  as $b$-chains of length $1$ over $\F_q^k$ and link them via the chains from Lemma~\ref{lemma_chain} to a $b$-chain with length $n_\lambda +\lambda'\cdot [k]_q$ having 
  the same start and end and being associated with a faithful projective $b-\big(n_\lambda +\lambda'\cdot [k]_q,k,\left(n_\lambda-d_\lambda\right)+\lambda'\cdot[k-b]_q\big)_q$
  system that corresponds to a $\left(n_\lambda+\lambda'\cdot[k]_q,k,d'+(\lambda+\lambda')\cdot q^{k-b}\cdot [b]_q\right)_q^b$ code. Using Lemma~\ref{lemma_griesmer_bound_period} 
  we conclude that the Griesmer bound is attained, i.e.\ the validity of the stated equation, for $d=d'+ (\lambda+\lambda')\cdot q^{k-b}\cdot [b]_q$. From 
  Proposition~\ref{prop_singer_construction}, Lemma~\ref{lemma_addition}, and Lemma~\ref{lemma_griesmer_bound_period} we then conclude that the Griesmer bound is 
  attained for all $d=d'+\lambda''\cdot q^{k-b}\cdot [b]_q$ where $\lambda''\ge \lambda+\lambda'$. 
\end{Proof}
Instead of decomposing $\cS_\lambda$ into chains of length $1$ we can also used decompositions into larger chains, that usually exist if the cardinality of $\cS_\lambda$ is 
not too small. We remark that the periodicity property of the Griesmer bound stated in Lemma~\ref{lemma_griesmer_bound_period} is accompanied by the upper bound
\begin{equation}
  n \le \frac{[k]_q\cdot s}{[k-b]_q}
\end{equation}  
for a projective $b-(n,k,s)_q$ system, that has an easy counting explanation, see e.g.\ \cite[Lemma 15]{kurz2024additive}. We remark that the Griesmer bound is always at least 
as good as this bound and attained with equality for the construction in Proposition~\ref{prop_singer_construction}. 

\medskip

The proof of Theorem~\ref{thm_main} in general gives constructions for rather large values of the minimum distance $d$ only. So, in order to find $[n,k,d]_q^b$ codes we can 
utilize ILP (integer linear programming) formulations. To this end let $\cP$ denote the set of points and $\cH$ denote the set of hyperplanes in $\PG(k-1,q)$. With this, let 
$\cT(b-1)$ be the elements in $\cP^{b-1}$ that span a $(b-1)$-space and $\cT(b)$ be the elements in $\cP^b$ that span a $b$-space.We use indicator variables $x_{S}^i\in \{0,1\}$ 
for all $S\in\cT(b)$ and all $0\le i\le n-1$. The interpretation is that $x_{\left(P_0,\dots,P_{b-1}\right)}^i=1$ iff the $(i+j)$th column of a generator matrix $G$ contains a 
representant of $P_j$ for all $0\le j\le b-1$. In order to ensure a unique choice we require
\begin{equation}
  \sum_{S\in\cT(b)} x_{S}^i =1\quad\forall 0\le i\le n-1.
\end{equation}
The \textit{chain property} is ensured via
\begin{eqnarray}
  &&\sum_{P_0\in\cP\,:\, \left(P_0,\dots,P_{b-1}\right)\in\cT(b)} \!\!\!\!\!\! x_{\left(P_0,\dots,P_{b-1}\right)}^i \notag\\ 
  &=& \sum_{P_b\in\cP\,:\,\left(P_1,\dots,P_{b}\right)\in\cT(b)} \!\!\!\!\!\! x_{\left(P_1,\dots,P_b\right)}^{i+1}
  \quad\forall 
  \left(P_1,\dots,P_{b-1}\right)\in\cT(b-1) 
  \,\forall 0\le i\le n-2
\end{eqnarray}
and
\begin{eqnarray}
  &&\sum_{P_0\in\cP\,:\, \left(P_0,\dots,P_{b-1}\right)\in\cT(b)} x_{\left(P_0,\dots,P_{b-1}\right)}^{n-1} \notag\\ 
  &=& \sum_{P_b\in\cP\,:\,\left(P_1,\dots,P_{b}\right)\in\cT(b)} x_{\left(P_1,\dots,P_b\right)}^{0}
  \quad\forall 
  \left(P_1,\dots,P_{b-1}\right)\in\cT(b-1). 
\end{eqnarray}
In order to count the maximum number of $b$-spaces per hyperplane we use
\begin{equation}
  \sum_{i=0}^{n-1} \sum_{\left(P_0,\dots,P_{b-1}\right)\in\cT(b) \,:\, \left\langle P_0,\dots,P_{b-1}\right\rangle \le H} x_{\left(P_0,\dots,P_{b-1}\right)}^i \le s \quad \forall H\in \cH. 
\end{equation} 
With this we either minimize $s$ and compute $d=n-s$ or we directly set $s=n-d$ and search for a feasible solution. We can add additional constraints mimicking our knowledge 
on the multiset of points covered by the $b$-spaces. I.e.\ we may e.g.\ prescribe precise point multiplicities or upper bounds. We may also assume 
that the generator matrix $G$ starts with an identity matrix.\footnote{By our definition of $\cT(b)$ and $\cT(b-1)$ we search for faithful codes only. Of course we can replace $\cT(b-1)$ 
by $\cP^{b-1}$ and $\cT(b)$ by $\cP^b$ for the more general situation. Note that we can assume that $G$ is obtained after an application of the Gauss-Jordan elimination algorithm, i.e.,
that $G$ is in reduced row echelon form.} We may also prescribe some automorphism $g$ and assume $x_{\left(P_0,\dots,P_{b-1}\right)}^i=x_{\left(P_0^g,\dots,P_{b-1}^g\right)}^{i+1}$ for all
$\left(P_0,\dots,P_{b-1}\right)\in\cT(b)$ and all $0\le i\le n-1$. 

\begin{Remark}
  Interpreting the sequence of columns of a generator matrix as a tour allows alternative formulations or relaxations. I.e. we may just use counting 
  variables $z_{\left(P_0,\dots,P_{b-1}\right)}:=\sum_{i=0}^{n-1} x_{\left(P_0,\dots,P_{b-1}\right)}^i\in \mathbb{N}$ instead of the $x_{\left(P_0,\dots,P_{b-1}\right)}^i\in\{0,1\}$. 
  Clearly we need 
  $$
    \sum_{P_0\in\cP\,:\,\left(P_0,\dots,P_{b-1}\right)\in\cT(b)} z_{\left(P_0,\dots,P_{b-1}\right)} =
    \sum_{P_b\in\cP\,:\,\left(P_1,\dots,P_{b}\right)\in\cT(b)} z_{\left(P_1,\dots,P_{b}\right)} 
  $$
  for all $\left(P_1,\dots,P_{b-1}\right)\in\cT(b-1)$. However, this formulation does note exclude the possibility that the desired tour is composed of several 
  subtours. So, we can use subtour elimination constraints as done by Dantzig, Fulkerson, and Johnson for the traveling salesperson problem. In the context of the latter
  optimization problem our formulation is similar to the idea used by Miller, Tucker, and Zemlin.   
\end{Remark}

\section{The functions $\mathbf{n_2^{2}(k,\cdot)}$ for $\mathbf{k\le 5}$} 
\label{sec_exact_values}
The aim of this section is to completely determine the function $n_2^{2}(k,\cdot)$ for the minimum possible length of an $[n,k,d]_2^2$ code for small dimensions $k$. Many 
of the presented methods are in principle also applicable for $[n,k,d]_q^b$ codes. However, our asymptotic result in Theorem~\ref{thm_main} usually applies for rather 
large values of $d$ only, so that many values of $n_q^{b}(k,d)$ would need to be determined to fully determine the function $n_q^b(k,\cdot)$ for other parameters.  

For each $c\in\F_q^n$ we obviously have $0\le \operatorname{wt}_2(c)\le n$ and $\operatorname{wt}_2(c)\in \mathbb{N}$. For small minimum distances $d$ the 
Singleton-type bound $\db(C)\le n+b-k$ \cite{ding2018maximum} is rather effective. More generally, the Griesmer-type bound, see Inequality~(\ref{eq_griesmer_bound_pair_distance}),  
can often by improved for small $d$ by applying Lemma~\ref{lemma_code_properties}, which is the source for many improved lower bounds for additive codes, see e.g.\ 
\cite{kurz2024additive}. Even for $q=b=2$ and $k\le 5$ there exist minimum distances $d$ where additive codes or the corresponding projective systems exist for some 
length $n$, but codes in the $b$-symbol metric require larger lengths, see e.g.\ Lemma~\ref{lemma_no_8_5_5} and Lemma~\ref{lemma_no_11_5_8}. For constructive upper 
bounds we mostly utilize the ILP formulation from the end of Section~\ref{sec_constructions}.   

For dimensions $k\le 2$ the determination of $n_q^b(k,\cdot)$ can still be solved completely:
\begin{Theorem}
  For each integer $d\ge 2$ we have $n_q^2(1,d)=n_q^2(2,d)=d$.
\end{Theorem}
\begin{Proof}
  Clearly we have $n_q^{2}(k,d)\ge d$ for all positive integers $k$ and $d$. For dimension $k=1$ we can consider a codeword $c\in\F_q^d$ consisting of $d$ ones 
  and check $\operatorname{wt}_2(\lambda\cdot c)=d$ for all $\lambda\in\F_q\backslash\{0\}$, so that $n_q^{2}(1,d)\le d$ for all $d\ge 2$. For dimension $k=2$ we consider 
  $$
    A=\begin{pmatrix}a_1\\a_2\end{pmatrix}=\begin{pmatrix} 1 & 0 \\ 0 & 1\end{pmatrix}\in\F_q^{2\times 2}\quad\text{and}\quad
    B=\begin{pmatrix}b_1\\b_2\end{pmatrix}=\begin{pmatrix} 1 & 1 & 0 \\ 0 & 1 & 1\end{pmatrix}\in\F_q^{2\times 3}.
  $$
  We easily check $\operatorname{d}_2(\operatorname{rowspan}(A))=2$ and $\operatorname{d}_2(\operatorname{rowspan}(B))=3$, so that Corollary~\ref{corollary_concatenation} implies  
  $\operatorname{d}_2(\operatorname{rowspan}(\overset{t\text{ times}}{\overbrace{A|\dots|A})})=2t$ 
  for all integers $t\ge 1$. In order to check $$\operatorname{d}_2(\operatorname{rowspan}(B|\overset{t\text{ times}}{\overbrace{A|\dots|A})})=
  \operatorname{d}_2(\operatorname{rowspan}(\overset{t\text{ times}}{\overbrace{A|\dots|A}}|B))=2t+3$$ we apply Lemma~\ref{lemma_concatenation} and Corollary~\ref{corollary_concatenation}.  
  More precisely, for some fixed integer $t\ge 1$ let 
  $c_1=(\overset{t\text{ times}}{\overbrace{a_1|\dots|a_1}})$ and $c_2=(\overset{t\text{ times}}{\overbrace{a_2|\dots|a_2}})$. For each $\lambda\in\F_q\backslash\{0\}$ both 
  $(\lambda\cdot b_1|\lambda\cdot c_1)$ and $(\lambda\cdot c_1|\lambda\cdot b_1)$ are of type (c).(iii) while $(\lambda\cdot b_2|\lambda\cdot c_2)$ and $(\lambda\cdot c_2|\lambda\cdot b_2)$ 
  are of type (b).(ii). For $\lambda_1,\lambda_2\in\F_q\backslash\{0\}$ we have that both $(\lambda_1b_1+\lambda_2b_2|\lambda_1c_1+\lambda_2c_2)$ and 
  $(\lambda_1c_1+\lambda_2c_2|\lambda_1b_1+\lambda_2b_2)$ are of type (d).(iv). Thus we have constructed examples showing $n_q^{2}(2,d)\le d$ for all integers $d\ge 2$.      
\end{Proof}

\begin{Theorem}
  For all integers $t\ge 0$ and $1\le i\le 6$ with $6t+i\ge 2$ we have $n_2^{2}(3,6t+i)= 7t+i+1$.  
\end{Theorem}
\begin{Proof}
  We choose 
  $$
    G_3=\begin{pmatrix}
    1 & 0 & 0 \\
    0 & 1 & 0 \\ 
    0 & 0 & 1
    \end{pmatrix}\!,\,\,
    G_4=\begin{pmatrix}
    1 & 0 & 0 & 1 \\
    0 & 1 & 0 & 1 \\ 
    0 & 0 & 1 & 1
    \end{pmatrix}\!,\,\,
    G_5=\begin{pmatrix}
    1 & 0 & 0 & 1 & 1 \\
    0 & 1 & 0 & 1 & 0 \\
    0 & 0 & 1 & 1 & 1
    \end{pmatrix}\!,\,\,
  $$
  $$
    G_6=\begin{pmatrix}
    1 & 0 & 0 & 1 & 1 & 1 \\ 
    0 & 1 & 0 & 1 & 0 & 1 \\
    0 & 0 & 1 & 0 & 1 & 1
    \end{pmatrix}\!,\,\,
    G_7=\begin{pmatrix}
    1 & 0 & 0 & 1 & 1 & 1 & 0 \\
    0 & 1 & 0 & 0 & 1 & 1 & 1 \\
    0 & 0 & 1 & 1 & 1 & 0 & 1
    \end{pmatrix}
  $$
  and verify $\operatorname{d}_2(\operatorname{rowspan}(G_i))=i-1$ for all $3\le i\le 7$ by exhaustively considering all corresponding seven non-zero codewords and computing the weights w.r.t.\ the 
  symbol-pair metric. From Lemma~\ref{lemma_addition} we conclude $n_2^2(3,6)\le n_2^2(3,3)+n_2^2(3,3)\le 8$ and $n_2^2(3,7)\le n_2^2(3,4)+n_2^2(3,3)\le 9$. For brevity, we denote
  corresponding generator matrices by $G_8$ and $G_9$, respectively.      
  Using Proposition~\ref{prop_singer_construction}, or Corollary~\ref{corollary_concatenation} applied to $G_7$. we conclude the 
  existence of $[7t,3,6t]_2^{2}$ codes for all positive integers $t$. Combining these with the codes generated by $G_3,\dots,G_9$ via Lemma~\ref{lemma_addition} yields the proposed lower bounds.  
  For the other direction we apply the Griesmer bound in Inequality~(\ref{eq_griesmer_bound_pair_distance}) and compute
  \begin{center} 
    \begin{tabular}{rrr}
      \hline 
      $i$ & $g_2(3,12t+2i)$ & $\left\lceil g_2(3,12t+2i)/3\right\rceil$\\ 
      \hline
      1 & $21t+4$  & $7t+2$ \\
      2 & $21t+7$  & $7t+3$ \\
      3 & $21t+11$ & $7t+4$ \\
      4 & $21t+14$ & $7t+5$ \\
      5 & $21t+18$ & $7t+6$ \\
      6 & $21t+21$ & $7t+7$ \\
      \hline
    \end{tabular}  
  \end{center}
\end{Proof}

We remark that examples of $[6,3,5]_2^{2}$ and $[7,3,6]_2^{2}$ codes can also be found in \cite[Table I]{chee2013maximum}, cf.~\cite[Table 1]{ding2018maximum}.  
Up to isomorphism there exists a unique $[n,3,n-1]_2^{2}$ code for all $n\in\{3,\dots,7\}$ while there exist $248$ non-isomorphic $[8,3,6]_2^{2}$ codes.

\begin{Corollary}
  We have $n_2^{2}(3,d)=\left\lceil\frac{g_2(3,2d)}{3}\right\rceil$ for all $d\ge 2$.
\end{Corollary}

\begin{Theorem}
  For all integers $t\ge 0$ and $1\le i\le 4$ with $4t+i\ge 5$ we have $n_2^{2}(4,4t+i)= 5t+i+1$. Moreover, we have $n_2^{2}(4,2)=4$, 
  $n_2^{2}(4,3)=5$, and $n_2^{2}(4,4)=6$.    
\end{Theorem}
\begin{Proof}
  Examples of $[4,4,2]_2^{2}$, $[5,4,3]_2^{2}$, $[6,4,4]_2^{2}$, $[7,4,5]_2^{2}$, $[8,4,6]_2^{2}$, $[9,4,7]_2^{2}$, $[10,4,8]_2^{2}$, 
  $[12,4,9]_2^{2}$, $[13,4,10]_2^{2}$, $[14,4,11]_2^{2}$, and $[15,4,12]_2^{2}$ codes are given by the generator matrices   
  $$
    \left(\begin{smallmatrix}
    1 0 0 0 \\ 
    0 1 0 0 \\ 
    0 0 1 0 \\ 
    0 0 0 1
    \end{smallmatrix}\right)\!,
    \left(\begin{smallmatrix}
    1 0 0 0 1 \\ 
    0 1 0 0 1 \\
    0 0 1 0 1 \\
    0 0 0 1 1
    \end{smallmatrix}\right)\!,
    \left(\begin{smallmatrix}
    1 0 0 0 1 0\\ 
    0 1 0 0 0 1\\ 
    0 0 1 0 1 0\\ 
    0 0 0 1 0 1
    \end{smallmatrix}\right)\!,
    \left(\begin{smallmatrix}
    1 0 0 0 1 1 0 \\ 
    0 1 0 0 0 1 1 \\
    0 0 1 0 1 1 1 \\
    0 0 0 1 1 0 1      
    \end{smallmatrix}\right)\!,
    \left(\begin{smallmatrix}
    1 0 0 0 1 1 1 0 \\ 
    0 1 0 0 1 0 1 1 \\
    0 0 1 0 0 1 1 1 \\
    0 0 0 1 1 1 0 1
    \end{smallmatrix}\right)\!,
    \left(\begin{smallmatrix}
    1 0 0 0 1 1 1 1 1 \\ 
    0 1 0 0 0 1 0 1 1 \\
    0 0 1 0 1 1 0 1 0 \\
    0 0 0 1 0 1 1 0 1
    \end{smallmatrix}\right)\!,
    \left(\begin{smallmatrix}
    1 0 0 0 1 1 1 1 1 1 \\ 
    0 1 0 0 1 0 1 0 1 0 \\
    0 0 1 0 0 1 1 0 1 1 \\
    0 0 0 1 1 0 1 1 0 1
    \end{smallmatrix}\right)\!,
  $$
  $$
    \left(\begin{smallmatrix}
    1 0 1 0 0 1 1 1 1 1 1 1 \\ 
    0 1 1 0 0 1 0 1 0 1 0 1 \\
    0 0 0 1 0 1 1 0 1 0 0 1 \\
    0 0 0 0 1 0 1 1 0 1 1 1     
    \end{smallmatrix}\right)\!, 
    \left(\begin{smallmatrix}
    1 0 1 0 0 1 1 1 1 1 1 1 1 \\
    0 1 1 0 0 0 1 1 0 1 0 1 0 \\
    0 0 0 1 0 1 0 1 1 0 0 1 1 \\
    0 0 0 0 1 1 1 1 0 1 1 0 1
    \end{smallmatrix}\right)\!,    
    \left(\begin{smallmatrix}
    1 0 1 0 0 1 1 1 1 1 1 0 0 0 \\ 
    0 1 1 0 0 1 0 1 0 0 1 1 1 1 \\
    0 0 0 1 0 1 1 1 0 1 0 1 1 0 \\
    0 0 0 0 1 0 1 1 1 0 1 1 0 1
    \end{smallmatrix}\right)\!,
    \left(\begin{smallmatrix}
    1 0 1 0 0 0 1 1 0 0 1 1 1 1 0 \\ 
    0 1 1 0 0 0 0 1 1 1 1 0 1 0 1 \\
    0 0 0 1 0 1 0 1 1 1 0 1 1 1 0 \\
    0 0 0 0 1 1 1 0 1 0 1 1 1 0 1
    \end{smallmatrix}\right)\!.
  $$
  By recursively applying Lemma~\ref{lemma_addition} we conclude $n_2^2(4,i+8t)\le i+2+10t$ for $5\le i\le 8$ as well as $n_2^2(4,i+8t)\le i+3+10t$ for $9\le i\le 12$, 
  where $t\in\mathbb{N}$ is arbitrary. Thus we have $n_2^{2}(4,4t+i)\le 5t+i+1$ for all integers $t\ge 0$ and $1\le i\le 4$ with $4t+i\ge 5$.     
  
  For the other direction we apply the Griesmer bound from Inequality~(\ref{eq_griesmer_bound_pair_distance}) and compute
  \begin{center} 
    \begin{tabular}{rrr}
      \hline 
      $i$ & $g_2(4,8t+2i)$ & $\left\lceil g_2(8,8t+2i)/3\right\rceil$\\ 
      \hline
      1 & $15t+5$  & $5t+2$ \\
      2 & $15t+8$  & $5t+3$ \\
      3 & $15t+12$ & $5t+4$ \\
      4 & $15t+15$ & $5t+5$ \\
      \hline
    \end{tabular}  
  \end{center}
  The improved lower bounds $n_2^{2}(4,2)\ge 4$, $n_2^{2}(4,3)\le 5$, and $n_2^{2}(4,4)\le 6$ are given by the  Singleton-type bound $d\le n+b-k$ \cite{ding2018maximum}. 
\end{Proof}
\begin{Corollary}
  We have $n_2^{2}(4,d)=\left\lceil\frac{g_2(4,2d)}{3}\right\rceil$ for all $d\ge 5$.
\end{Corollary}

\begin{table}[htp]
  \begin{center}
    \begin{tabular}{|r|rrrrrrrrrrr|}
      \hline
      $d$  & 2 & 3 & 4 & 5 & 6 & 7 & 8 & 9 & 10 & 11 & 12 \\
      \hline
      $n_2^{2}(4,d)$ & 4 & 5 & 6 & 7 & 8 & 9 & 10 & 12 & 13 & 14 & 15 \\  
      \hline 
      $\#$ & 1 & 1 & 4 & 1 & 4 & 1 & 1 &  21030 &  13772 & 755 & 33 \\
      \hline
    \end{tabular}
    \caption{Number of non-isomorphic optimal $\left[n_2^{2}(4,d),4,d\right]_2^{2}$ codes}
  \end{center}
\end{table}

\begin{Theorem}
  \label{thm_exact_5}
  We have $n_2^{2}(5,d)=\left\lceil\frac{g_2(5,2d)}{3}\right\rceil$ for all $d\ge 9$. 
  Moreover, we have $n_2^{2}(5,d)=d+3$ for $d\in\{2,3,4,6,7\}$, $n_2^{2}(5,5)=9$, and $n_2^{2}(5,8)=12$.    
\end{Theorem}
\begin{Proof}
  Examples of $[5,5,2]_2^{2}$, $[6,5,3]_2^{2}$, $[7,5,4]_2^{2}$, $[9,5,5]_2^{2}$, $[9,5,6]_2^{2}$, $[10,5,7]_2^{2}$, $[12,5,8]_2^{2}$, 
  $[13,5,9]_2^{2}$, $[14,5,10]_2^{2}$, $[15,5,11]_2^{2}$, $[16,5,12]_2^{2}$, $[18,5,13]_2^{2}$, $[19,5,14]_2^{2}$, $[20,5,15]_2^{2}$,
  $[21,5,16]_2^2$, $[23,5,17]_2^{2}$, $[24,5,18]_2^{2}$, $[25,5,19]_2^{2}$, $[26,5,20]_2^{2}$, $[28,5,21]_2^{2}$, $[29,5,22]_2^{2}$, 
  $[30,5,23]_2^{2}$, $[31,5,24]_2^{2}$, $[33,5,25]_2^{2}$, $[34,5,26]_2^{2}$, $[36,5,27]_2^{2}$, $[37,5,28]_2^{2}$, $[38,5,29]_2^{2}$, 
  $[39,5,30]_2^{2}$, $[41,5,31]_2^{2}$, and $[42,5,32]_2^{2}$ codes are given by the generator matrices   
  $$
    \left(\begin{smallmatrix}
    1 0 0 0 0 \\ 
    0 1 0 0 0 \\ 
    0 0 1 0 0 \\ 
    0 0 0 1 0 \\ 
    0 0 0 0 1
    \end{smallmatrix}\right)\!,
    \left(\begin{smallmatrix}
    1 0 0 0 0 1 \\
    0 1 0 0 0 1 \\
    0 0 1 0 0 1 \\
    0 0 0 1 0 1 \\
    0 0 0 0 1 1
    \end{smallmatrix}\right)\!,
    \left(\begin{smallmatrix}
    1 0 0 0 0 1 0 \\
    0 1 0 0 0 1 1 \\
    0 0 1 0 0 0 1 \\
    0 0 0 1 0 1 1 \\
    0 0 0 0 1 0 1 
    \end{smallmatrix}\right)\!,
    \left(\begin{smallmatrix}
    1 1 0 0 0 0 1 1 1 \\ 
    0 0 1 0 0 0 1 1 0 \\
    0 0 0 1 0 0 1 1 1 \\
    0 0 0 0 1 0 1 0 1 \\
    0 0 0 0 0 1 0 1 1
    \end{smallmatrix}\right)\!,
    \left(\begin{smallmatrix}
    1 0 0 0 0 1 1 1 1 \\
    0 1 0 0 0 1 0 0 1 \\
    0 0 1 0 0 1 0 1 0 \\
    0 0 0 1 0 1 1 0 1 \\
    0 0 0 0 1 1 0 1 1
    \end{smallmatrix}\right)\!,
    \left(\begin{smallmatrix}
    1 0 0 0 0 1 1 1 1 0 \\ 
    0 1 0 0 0 0 1 0 1 1 \\
    0 0 1 0 0 1 1 0 0 1 \\
    0 0 0 1 0 1 0 1 1 1 \\
    0 0 0 0 1 0 1 1 0 1
    \end{smallmatrix}\right)\!,
    \left(\begin{smallmatrix}
    1 1 0 0 0 0 1 1 1 1 1 0 \\ 
    0 0 1 0 0 0 0 1 1 1 0 1 \\
    0 0 0 1 0 0 1 0 1 0 1 0 \\
    0 0 0 0 1 0 1 1 1 0 0 1 \\
    0 0 0 0 0 1 1 0 1 1 1 1    
    \end{smallmatrix}\right)\!,
  $$
  $$
    \left(\begin{smallmatrix}
    1 0 0 1 0 0 1 1 1 1 1 1 1 \\ 
    0 1 0 1 0 0 1 0 1 0 1 0 1 \\
    0 0 1 1 0 0 0 1 1 0 0 1 1 \\
    0 0 0 0 1 0 1 1 0 1 1 0 1 \\
    0 0 0 0 0 1 1 0 1 1 0 1 1
    \end{smallmatrix}\right)\!,
    \left(\begin{smallmatrix}
    1 0 0 1 0 0 1 1 1 1 1 1 1 1 \\ 
    0 1 0 1 0 0 1 0 1 0 0 0 0 1 \\
    0 0 1 1 0 0 0 1 0 0 1 0 1 1 \\
    0 0 0 0 1 0 1 1 0 1 1 0 0 1 \\
    0 0 0 0 0 1 1 0 1 0 1 1 1 1
    \end{smallmatrix}\right)\!,
    \left(\begin{smallmatrix}
    1 1 1 0 0 0 0 1 1 0 1 1 1 1 0 \\
    1 1 1 0 1 1 1 0 0 0 0 1 0 1 1 \\
    1 1 1 1 0 0 1 1 0 1 1 0 0 0 1 \\
    1 1 0 0 0 1 1 0 1 1 1 0 0 1 0 \\
    0 1 1 0 0 1 0 0 0 1 1 0 1 1 1
    \end{smallmatrix}\right)\!,
    \left(\begin{smallmatrix}
    0 1 0 1 0 1 0 1 0 1 1 0 1 1 0 1 \\ 
    0 0 0 0 0 1 0 1 1 0 1 1 1 0 1 1 \\
    0 0 0 0 1 0 1 0 0 1 0 1 0 1 0 1 \\
    0 0 1 1 0 0 1 1 0 0 0 1 1 1 1 1 \\
    1 0 0 1 0 0 1 0 0 0 1 0 1 0 1 0
    \end{smallmatrix}\right)\!,
    \left(\begin{smallmatrix}
    0 0 0 1 0 1 1 0 1 0 1 0 1 1 0 1 0 1 \\ 
    0 0 1 0 0 0 0 1 1 0 1 1 0 1 1 1 1 1 \\
    0 1 0 0 0 0 1 0 0 1 0 1 1 0 1 1 0 1 \\
    0 0 0 0 1 0 0 0 1 1 0 1 1 1 1 0 1 1 \\
    1 0 0 0 0 1 0 1 0 1 1 1 1 1 0 0 0 1
    \end{smallmatrix}\right)\!, 
  $$  
  $$
    \left(\begin{smallmatrix}
    0 0 0 0 1 1 0 1 1 1 0 1 1 0 1 1 0 1 1 \\
    0 0 1 1 0 1 0 0 0 1 1 0 1 1 0 1 1 0 1 \\
    0 0 0 1 1 1 1 0 1 0 0 0 0 1 0 1 1 1 1 \\
    1 1 0 0 0 0 0 1 1 1 1 0 0 1 1 1 1 0 1 \\ 
    0 1 0 0 1 0 1 1 1 0 0 1 0 1 0 0 0 0 1
    \end{smallmatrix}\right)\!, 
    \left(\begin{smallmatrix}
    0 0 0 0 0 1 1 0 1 1 1 0 1 1 0 1 0 1 0 1 \\ 
    0 0 0 1 1 0 1 1 0 1 1 1 0 1 1 0 0 1 1 0 \\
    0 1 1 0 0 0 0 1 0 0 1 1 0 1 0 1 1 0 1 1 \\
    0 0 1 0 1 0 1 0 1 0 0 0 1 0 0 1 1 0 1 0 \\
    1 0 0 0 0 0 1 0 0 1 0 1 1 1 1 0 1 0 1 1
    \end{smallmatrix}\right)\!, 
    \left(\begin{smallmatrix}
    0 0 0 0 0 1 1 0 1 0 1 1 1 0 1 1 0 1 1 0 1 \\ 
    0 0 0 1 1 0 0 0 1 1 0 1 1 1 0 1 1 0 1 1 0 \\ 
    0 0 1 0 1 0 1 1 0 0 1 0 1 1 1 0 0 0 1 0 1 \\
    1 1 0 0 0 0 0 1 0 1 0 1 0 1 1 1 0 0 1 1 1 \\
    0 1 0 0 1 0 1 0 1 1 0 1 0 0 0 0 1 1 0 0 1
    \end{smallmatrix}\right)\!,
    \left(\begin{smallmatrix}
    0 0 0 0 0 0 1 1 1 0 1 1 1 0 1 1 0 1 1 1 0 1 1 \\ 
    0 0 0 0 1 1 0 0 1 1 0 1 1 1 0 1 1 0 0 1 1 0 1 \\ 
    0 0 1 1 0 1 0 1 1 1 1 0 1 1 0 0 0 0 1 1 0 1 0 \\
    0 1 0 1 0 0 0 0 0 1 1 1 1 0 0 1 1 1 0 1 0 1 0 \\ 
    1 0 0 1 0 0 0 1 1 0 1 1 0 1 1 0 0 0 0 1 1 0 1
    \end{smallmatrix}\right)\!,
  $$
  $$
    \left(\begin{smallmatrix}
    1 0 1 0 0 0 1 0 1 0 0 0 1 0 1 1 0 1 1 1 1 0 1 1 \\ 
    0 1 1 0 0 0 0 1 1 0 1 1 0 1 1 0 0 0 1 0 0 1 1 1 \\ 
    0 0 0 1 0 0 1 0 0 1 0 1 0 1 0 1 1 1 0 0 1 0 1 1 \\
    0 0 0 0 1 0 1 1 1 0 1 0 1 0 0 1 1 0 1 1 0 0 0 1 \\
    0 0 0 0 0 1 1 1 0 1 0 1 1 0 1 0 1 1 1 0 0 1 1 0
    \end{smallmatrix}\right)\!,
    \left(\begin{smallmatrix}
    0 1 1 1 1 1 0 1 1 0 1 0 1 0 0 0 0 1 1 0 1 1 0 0 0 \\ 
    0 1 0 1 1 1 1 0 0 1 0 0 1 1 1 0 1 0 0 1 1 0 1 0 0 \\
    0 0 1 1 1 1 1 0 1 0 1 1 0 1 1 1 1 1 0 0 0 0 0 1 0 \\
    0 1 0 0 1 1 1 1 1 1 1 1 0 1 0 0 0 0 1 1 1 0 0 0 1 \\
    1 1 1 0 1 0 1 0 1 0 0 1 1 0 1 1 0 0 1 1 0 0 0 0 0
    \end{smallmatrix}\right)\!,
    \left(\begin{smallmatrix} 
    0 1 1 0 1 1 1 1 1 1 1 0 1 0 1 0 0 1 1 0 0 0 1 0 0 0 \\ 
    0 1 0 0 1 1 0 1 0 0 0 1 0 1 1 1 0 0 1 0 1 1 0 1 0 0 \\
    0 1 0 1 1 1 1 0 1 0 1 1 0 0 0 0 0 1 0 1 0 1 0 0 1 0 \\
    0 0 1 1 1 1 0 0 1 1 1 0 0 0 1 1 1 0 1 1 1 1 0 0 0 1 \\
    1 1 1 1 1 0 0 1 0 0 1 1 1 1 0 1 1 1 1 0 0 1 0 0 0 0
    \end{smallmatrix}\right)\!,
  $$
  $$  
    \left(\begin{smallmatrix}
    0 1 1 1 1 0 1 1 0 1 1 0 1 0 1 0 1 0 0 1 0 0 0 0 1 0 0 0 \\ 
    0 1 0 1 1 1 1 0 1 1 0 0 0 0 1 1 1 0 0 0 1 1 1 1 0 1 0 0 \\
    0 0 0 0 0 1 1 1 0 1 0 1 0 0 1 0 1 1 1 1 1 0 1 1 0 0 1 0 \\
    0 1 1 0 1 0 1 0 0 0 1 0 0 1 0 1 1 1 1 1 1 1 1 0 0 0 0 1 \\
    1 0 0 0 1 0 0 1 1 1 1 1 1 1 0 0 1 0 1 1 0 1 1 1 0 0 0 0
    \end{smallmatrix}\right)\!,
    \left(\begin{smallmatrix}
    0 1 1 1 1 1 0 0 0 0 1 1 1 0 1 1 1 1 0 1 0 1 0 0 1 1 0 0 0 \\
    0 1 0 1 0 0 1 1 0 1 0 1 1 0 0 1 0 1 0 1 1 1 1 1 0 0 1 0 0 \\
    0 1 0 0 1 0 0 1 1 1 1 1 1 1 0 0 1 0 1 1 1 0 1 0 1 0 0 1 0 \\
    0 0 0 1 0 1 0 0 1 1 1 0 1 1 1 0 1 0 0 1 1 1 0 1 0 0 0 0 1 \\
    1 0 1 1 1 0 1 0 1 1 0 1 0 0 1 0 1 1 1 1 0 0 1 0 0 0 0 0 0
    \end{smallmatrix}\right)\!,  
    \left(\begin{smallmatrix}
    0 1 1 1 0 0 1 0 0 1 1 0 0 1 0 1 1 0 1 1 0 0 1 0 1 1 1 0 0 0 \\ 
    0 1 0 0 0 1 1 1 0 1 1 1 1 0 1 0 1 1 0 1 0 1 0 0 1 0 0 1 0 0 \\ 
    0 1 0 1 1 1 1 0 0 0 1 0 0 1 1 1 0 1 0 0 1 1 0 1 0 1 0 0 1 0 \\
    0 0 1 1 1 1 0 0 1 0 1 1 1 1 0 0 1 0 1 1 0 1 0 1 0 0 0 0 0 1 \\
    1 1 1 1 0 1 0 1 1 1 0 0 1 0 1 0 1 0 0 0 1 0 1 1 0 1 0 0 0 0
    \end{smallmatrix}\right)\!, 
  $$  
  $$
    \left(\begin{smallmatrix}
    0 1 1 0 0 1 0 0 1 1 1 1 1 0 1 1 1 0 0 0 1 0 1 0 1 1 0 1 0 0 0 \\ 
    0 0 1 1 0 0 1 0 0 1 1 1 1 1 0 1 1 1 0 0 0 1 0 1 0 1 1 0 1 0 0 \\
    0 1 1 1 1 1 0 1 1 1 0 0 0 1 0 1 0 1 1 0 1 0 0 0 0 1 1 0 0 1 0 \\
    0 1 0 1 1 0 1 0 0 0 0 1 1 0 0 1 0 0 1 1 1 1 1 0 1 1 1 0 0 0 1 \\
    1 1 0 0 1 0 0 1 1 1 1 1 0 1 1 1 0 0 0 1 0 1 0 1 1 0 1 0 0 0 0
    \end{smallmatrix}\right)\!,
    \left(\begin{smallmatrix}
    1 0 0 0 0 1 1 1 0 0 1 0 1 1 0 1 0 1 1 0 0 0 1 1 0 0 1 1 1 1 0 0 1 \\ 
    0 1 0 0 0 1 0 1 1 1 1 0 1 0 1 1 0 0 0 1 0 0 1 0 1 0 0 0 1 0 0 1 1 \\
    0 0 1 0 0 1 0 1 0 1 1 1 1 1 0 0 0 1 0 1 1 0 0 1 1 0 1 0 1 1 1 0 0 \\
    0 0 0 1 0 0 1 1 0 1 1 1 0 1 1 1 0 1 1 1 1 1 0 0 0 1 0 0 1 0 0 0 1 \\
    0 0 0 0 1 0 1 0 0 0 1 1 0 0 1 0 1 1 0 1 0 0 0 1 0 1 0 1 1 1 1 1 1
    \end{smallmatrix}\right)\!,
  $$
  $$
    \left(\begin{smallmatrix}
    1 0 0 0 0 1 0 0 1 0 1 0 1 1 0 0 1 1 1 1 0 1 1 1 0 0 1 0 1 0 1 1 1 0 \\ 
    0 1 0 0 0 0 1 1 1 1 0 1 0 1 0 1 0 0 1 1 0 0 1 0 1 0 1 1 0 0 0 1 1 0 \\
    0 0 1 0 0 0 0 0 1 1 1 0 0 0 0 1 0 1 0 1 1 0 1 1 1 0 1 0 1 1 1 0 1 1 \\
    0 0 0 1 0 1 1 0 0 0 0 1 0 1 1 1 1 1 0 0 1 0 1 0 1 1 0 1 1 1 0 0 1 0 \\
    0 0 0 0 1 0 0 1 0 0 1 0 1 0 1 0 1 1 0 1 1 1 1 0 1 0 0 1 0 0 0 1 0 1
    \end{smallmatrix}\right)\!,
    \left(\begin{smallmatrix}
    1 0 0 0 0 1 1 0 0 0 0 0 1 1 0 1 0 1 0 1 0 1 0 1 1 0 1 1 0 1 1 1 0 0 0 1 \\ 
    0 1 0 0 0 0 1 1 1 0 1 1 1 0 1 1 0 1 1 1 1 0 0 1 0 0 0 1 0 1 1 0 1 0 0 0 \\
    0 0 1 0 0 1 1 1 0 0 0 1 1 0 1 1 1 0 0 0 1 1 1 1 1 0 1 0 0 1 0 0 1 0 1 1 \\
    0 0 0 1 0 0 1 0 0 1 0 0 0 1 1 1 1 0 1 1 1 1 0 0 0 1 1 1 0 0 1 0 1 1 1 1 \\
    0 0 0 0 1 1 0 1 1 1 1 1 0 0 1 1 1 1 1 0 0 0 1 1 1 0 0 0 1 0 1 0 1 1 0 1
    \end{smallmatrix}\right)\!, 
  $$
  $$
    \left(\begin{smallmatrix}
    1 0 0 0 0 1 0 0 1 1 1 1 1 0 1 1 0 1 0 1 0 1 0 1 0 0 1 0 0 1 1 1 1 0 0 0 0 \\ 
    0 1 0 0 0 1 1 0 0 1 0 0 1 0 1 0 0 0 0 0 0 1 1 1 0 1 0 1 1 0 1 0 1 1 0 1 1 \\
    0 0 1 0 0 1 0 1 1 0 1 1 1 1 0 0 1 0 1 0 0 0 0 1 0 1 1 1 0 0 1 0 1 0 0 1 0 \\
    0 0 0 1 0 1 1 1 0 0 1 0 0 1 1 1 0 0 1 1 0 0 0 1 1 1 0 1 1 0 0 0 1 0 1 0 1 \\
    0 0 0 0 1 1 1 0 0 1 1 1 0 1 0 1 1 1 1 0 1 0 1 0 0 1 0 0 0 1 1 0 1 0 1 0 0
    \end{smallmatrix}\right)\!,
    \left(\begin{smallmatrix}
    1 0 0 0 0 1 0 0 1 1 0 1 1 1 1 0 1 1 1 1 0 0 1 1 0 0 1 1 1 0 0 0 0 1 1 0 1 1 \\ 
    0 1 0 0 0 1 0 0 1 1 1 0 1 0 0 1 0 0 0 0 1 1 1 0 1 1 1 1 1 1 0 1 1 0 1 1 1 0 \\
    0 0 1 0 0 1 0 1 0 1 1 1 1 1 1 1 0 1 0 1 1 0 0 0 0 1 0 1 1 1 1 1 1 1 1 0 0 1 \\
    0 0 0 1 0 1 1 1 0 0 1 0 1 0 0 1 1 1 1 1 0 1 0 0 1 0 0 0 1 1 1 1 0 1 0 0 1 1 \\
    0 0 0 0 1 0 1 0 1 0 1 1 1 0 1 1 0 1 0 0 1 1 0 1 0 1 1 1 0 0 1 0 0 0 0 0 1 1
    \end{smallmatrix}\right)\!,
  $$
  $$
    \left(\begin{smallmatrix}
    1 0 0 0 0 0 1 0 0 1 0 0 1 0 0 0 1 0 1 0 1 1 1 1 0 1 1 0 1 1 1 1 0 0 1 0 1 1 0 \\
    0 1 0 0 0 1 0 1 0 1 0 0 0 1 0 1 1 1 1 0 1 1 0 1 1 0 1 1 1 0 0 0 1 1 0 0 1 0 0 \\
    0 0 1 0 0 1 1 1 1 1 1 1 0 0 0 0 0 0 1 0 0 1 0 0 1 0 1 1 1 1 1 1 0 1 1 1 1 0 1 \\
    0 0 0 1 0 0 0 1 1 0 0 1 1 1 1 1 0 1 1 0 1 1 1 1 0 0 1 1 0 1 0 1 0 1 0 0 0 0 1 \\ 
    0 0 0 0 1 0 1 1 0 0 1 1 1 1 1 1 1 0 0 1 0 0 0 1 1 1 1 0 1 1 1 0 1 1 0 1 1 0 1
    \end{smallmatrix}\right)\!,
    \left(\begin{smallmatrix}
    1 0 0 0 0 1 0 0 1 0 0 0 1 1 1 1 1 0 0 0 0 1 0 1 0 1 0 1 1 1 1 0 1 0 1 1 0 0 1 0 1 \\ 
    0 1 0 0 0 1 1 1 1 0 1 1 1 0 0 1 0 1 0 0 0 0 0 1 0 1 0 0 1 0 0 1 0 1 1 0 1 1 1 1 0 \\
    0 0 1 0 0 0 1 0 0 1 1 0 0 1 0 1 0 0 1 1 0 0 1 0 0 1 0 1 0 0 1 0 1 1 1 1 0 1 0 1 0 \\
    0 0 0 1 0 1 0 0 0 0 1 1 0 0 0 1 1 0 1 1 1 0 1 0 1 0 0 1 0 1 0 0 1 0 1 0 1 1 1 1 1 \\
    0 0 0 0 1 0 0 1 0 0 1 0 0 0 0 0 1 1 1 0 1 1 0 1 0 1 1 0 1 0 1 0 1 1 1 1 1 0 1 1 1
    \end{smallmatrix}\right)\!, 
  $$
  $$
    \left(\begin{smallmatrix}
    1 0 0 0 0 0 0 0 0 1 0 1 1 1 0 1 1 0 1 0 1 0 1 1 0 0 0 1 0 0 1 0 1 1 1 1 0 1 0 1 1 0 \\ 
    0 1 0 0 0 1 0 0 1 1 1 0 0 1 0 0 1 0 0 0 0 1 1 1 0 1 0 1 1 1 0 1 1 1 1 0 1 0 1 1 0 0 \\
    0 0 1 0 0 1 0 0 0 1 0 1 0 1 1 0 1 1 1 0 1 0 0 1 1 0 0 0 0 1 1 1 1 0 1 0 1 0 0 0 1 1 \\
    0 0 0 1 0 1 1 0 1 0 1 0 0 1 1 1 0 0 0 1 0 0 1 0 1 0 1 0 1 0 1 0 1 0 1 0 1 1 0 0 0 0 \\
    0 0 0 0 1 0 1 1 0 1 0 1 0 1 0 0 1 0 1 1 0 0 0 0 1 1 0 1 1 0 0 1 0 0 1 1 1 1 1 0 0 1
    \end{smallmatrix}\right)\!.
  $$
  (A $[41,5,31]_2^{2}$ code can also be constructed from a $[31,5,24]_2^{2}$ and a $[10,5,7]_2^{2}$ code.  
  For a general construction of a $[31,5,24]_2^2$ code we refer to Proposition~\ref{prop_singer_construction}.) 
  Combining these codes with a suitable number of $[31,5,24]_2^2$ codes via Lemma~\ref{lemma_addition} gives the proposed 
  upper bounds for $n_2^2(5,d)$ for all $d\ge 33$.   
  
  The proposed lower bounds for $n_2^2(5,d)$ are given by the Griesmer bound from Inequality~(\ref{eq_griesmer_bound_pair_distance}) for all $d\ge 9$ 
  and by the Singleton-type bound $d\le n+b-k$ \cite{ding2018maximum} for all $d\in\{2,3,4,6,7\}$. 
  Using \texttt{LinCode} we have enumerated all $10358$ even $[24,5,10]_2$ codes. While several of the corresponding multisets of points can be partitioned into lines, no partition 
  yields a $[8,5,5]_2^2$ code, so that $n_2^2(5,5)\ge 9$. Alternatively, we can directly use Lemma~\ref{lemma_no_8_5_5}. Using \texttt{LinCode} we have enumerated 
  all $21$ even $[33,5,16]_2$ codes. While several of the corresponding multisets of points can be partitioned into lines, no partition yields a $[11,5,8]_2^2$ code, so 
  that $n_2^2(5,8)\ge 12$. Alternatively, we can directly use Lemma~\ref{lemma_no_11_5_8}. 
\end{Proof} 

We remark that in \cite[Table I]{chee2013maximum} generator matrices for $[7,4,5]_2^{2}$, $[8,4,6]_2^{2}$, $[9,4,7]_2^{2}$, and $[10,4,8]_2^{2}$ codes are 
stated. The length of the list of explicitly constructed generator matrices is partially due to the fact that the existence of an $[n,k,d]_q^b$ code does
not necessarily imply the existence of an $[n-1,k,d-1]_q^b$ code, see Example~\ref{example_singer}.

Our next aim is to replace the computer enumerations for $d\in \{5,8\}$ in the proof of Theorem~\ref{thm_exact_5} by theoretical arguments.
\begin{Lemma}
  \label{lemma_multiset}
  Given an $[n,k,d]_q^{2}$ code $C$ with $n=n_q^2(k,d)$, there exists a spanning multiset of points in $\PG(k-1,q)$ of cardinality $(q+1)n$ that can be written as 
  $\cM=\sum_{i=0}^{n-1} \chi_{L_i}$ for $n$ lines $L_0,\dots,L_{n-1}$ such that each hyperplane $H$ contains at most $(n-d)$ of the lines and $\sum_{P\le H} \mathcal{M}(P)\le q(q+1)-dq$. 
  If $q=2$ there additionally exist $n$ points $P_0,\dots P_{n-1}$ such that $\cM=\sum_{0}^{n-1} \left(2\cdot\chi_{P_i}+\chi_{Q_i}\right)$, where $Q_i=P_i+P_{i+1}$ (reading indices modulo $n$).
\end{Lemma}
\begin{Proof}
  Due to Lemma~\ref{lemma_faitful} we can assume that $C$ is faithful. Given a generator matrix $G$ of $C$ we denote the spans of the columns of 
  $G$ by $P_0,\dots,P_{n-1}$ and set $L_i:=\left\langle P_i,P_{i+1}\right\rangle$, where the indices are read modulo $n$. Setting $\mathcal{M}:=\sum_{i=0}^{n-1} \chi_{L_i}$ 
  gives the multiset of points associated to $C$. From Lemma~\ref{lemma_associated_projective_system} we conclude that at most $n-d$ lines can be fully contained in a hyperplane.    
  Since each hyperplane either fully contains a line  or intersects it in a point we conclude the upper bound $\sum_{P\le H} \mathcal{M}(P)\le q(q+1)-dq$ for all hyperplanes $H$. 
  For $q=2$ we observe $L_i=\left\{P_i,Q_i,P_{i+1}\right\}$ for all $0\le i\le n-1$.
\end{Proof}

\begin{Lemma}
  \label{lemma_no_8_5_5}
  No $[8,5,5]_2^{2}$ code exists.
\end{Lemma}
\begin{Proof}
  Assuming that such a code exist we use Lemma~\ref{lemma_multiset} to construct a multiset of points $\mathcal{M}=\sum_{i=0}^7 \left(2\chi_{P_i}+\chi_{Q_i}\right)$, where $Q_i=P_i+P_{i+1}$ 
  with indices read modulo $8$, 
  that corresponds to a $[24,5,10]_2$ code using Lemma~\ref{lemma_multiset}. Since the points $P_0,\dots,P_7$ span $\PG(4,2)$ and each $[8,5,d]_2$ code satisfies $d\le 2$ we conclude the 
  existence of a hyperplane $H$ that contains at least six of the points $P_0,\dots,P_7$. Thus, $H$ contains at least four of the points $Q_0,\dots,Q_7$, so that the multiplicity of $H$ 
  is at least $2\cdot 6+4=16$. However, the maximum possible multiplicity of a hyperplane in a multiset of points corresponding to a $[24,5,10]_2$ code is $24-10=14$ -- contradiction.  
\end{Proof}

In order to show the non-existence of a $[11,5,8]_2^{2}$ code we need to refine the argument a bit. (We may deduce the existence of a hyperplane $H$ that contains a least seven of the 
eleven double points, say $\left\{P_0,P_1,P_3,P_4,P_6,P_7,P_9\right\}$ while the points in $\left\{P_2,P_5,P_8,P_{10}\right\}$ are not contained. In this example only three of the eleven points 
$Q_i$ are contained in $H$, which gives a multiplicity of $2\cdot 7+3=17$ for $H$, which goes in line with a $[33,5,16]_2$ code.)

\begin{Lemma}
  \label{lemma_no_11_5_8}
  No $[11,5,8]_2^{2}$ code exists.
\end{Lemma}
\begin{Proof}
  Assuming that such a code exist we construct a multiset of points $\mathcal{M}=\sum_{i=0}^{10} \chi_{L_i}$ as in Lemma~\ref{lemma_multiset} corresponding to a $[33,5,16]_2$ code, where the $L_i$ 
  are lines. By construction we have $\mathcal{M}(L_i)\ge 2\cdot 2+1=5$. Since no $[27,3,16]_2$ code exists we have indeed $\mathcal{M}(L_i)=5$ for all $0\le i\le 10$. Projection through 
  $L_i$ yields a multiset of points $\mathcal{M}'$ that corresponds to a $[28,3,16]_2$ code. Since we  have $\mathcal{M}'(L')\le 12$ for every line $L'$ in $\PG(4,2)/L_i\cong \PG(2,2)$ 
  we conclude $\mathcal{M}'(P')\le 4$ for every point $P'$ in the factor space. Since there are only seven points and $\mathcal{M}'$ has cardinality $28$ we have $\mathcal{M}'(P')=4$ for all points. 
  Thus, we have $\mathcal{M}(H)=5+3\cdot 4=17$ for every hyperplane $H\ge L_i$. Counting gives that those hyperplanes contain exactly three lines. Since we can choose $0\le i\le 10$ arbitrarily we 
  can state that each hyperplane that contains at least one of the eleven lines contains exactly three of them. Denoting the number of hyperplanes that contain at least one line by $x$ and 
  double counting lines gives
  $$
    3\cdot x + 0\cdot(31-x)=7\cdot 11,
  $$    
  which does not have an integral solution -- contradiction.
\end{Proof}

\section{Conclusion}
\label{sec_conclusion}

In Theorem~\ref{thm_main} we have shown that the minimum possible length $n_q^b(k,d)$ of an $[n,k,d]_q^b$ code is attained by a Griesmer-type bound 
$n_q^b(k,d)\ge \left\lceil \frac{g_q\!\left(k,q^{b-1}\cdot d\right)}{[b]_q} \right\rceil$ if $d$ is sufficiently large. With this the problem of the 
determination of $n_q^b(k,\cdot)$ becomes a finite problem. In Section~\ref{sec_exact_values} we have solved this problem for $q=b=2$ and $k\le 5$. Besides 
the general construction in Proposition~\ref{prop_singer_construction} and Lemma~\ref{lemma_addition} for the combination of codes, we only used explicit codes 
found by ILP searches. In order to determine the functions $n_q^b(k,\cdot)$ for more parameters, more general constructions are desired. Although we have 
described $[n,k,d]_q^b$ codes from the geometric point of view as projective $b-(n,k,n-d)_q$ systems we are still very far from a one-to-one correspondence.   


\end{document}